%% file: Kalman_main.tex
\documentclass[journal]{IEEEtran}
\ifCLASSINFOpdf
\else
\fi
\usepackage{colortbl}
\usepackage[ruled,linesnumbered]{algorithm2e} 
\usepackage{graphicx}
\usepackage{textcomp}
\usepackage{xcolor}
\usepackage[nolist,nohyperlinks]{acronym}
\usepackage{lipsum}  
\usepackage{float}
\usepackage{bm}
\usepackage{flushend}
\usepackage{cuted}
\usepackage{needspace}
\usepackage{rotating}
\usepackage{bbm}
\usepackage[utf8]{inputenc}
\usepackage{tikz}
\usepackage{tablefootnote}
\usepackage[inline]{enumitem}
\usepackage{multirow}
\usepackage{array}
\usepackage{colortbl}
\usepackage{verbatim}
\usepackage{url}
\usepackage{soul}
\usepackage{amsmath,amsfonts,amssymb}
\usepackage{subcaption} 
\usepackage{enumitem}

\usepackage[utf8]{inputenc}
\usepackage[english]{babel}
\usepackage[sort,square,numbers]{natbib}

\newcommand{\sectionlabel}{Section }

\newcommand{\predlength}[1]{$L = #1$}
\newcommand{\filterwindow}[1]{$F = #1$}

\begin{document}
%
\title{A Kalman Filter based Low Complexity Throughput Prediction Algorithm for 5G Cellular Networks}
%
%
%

\author{Mayukh Biswas, 
        Ayan Chakraborty, and Basabdatta Palit
\thanks{M. Biswas is with the Department
of Electronics and Telecommunication Engineering, IIEST Shibpur, Howrah - 71103, West Bengal, India.}
\thanks{A. Chakraborty is with the G.S.Sanyal School of Telecommunications, IIT Kharagpur, Kharapur - 721302, West Bengal, India}
\thanks{B.Palit is with the Department
of Electronics and Communication Engineering, NIT Rourkela, Rourkela - 769008, Odisha, India.}
}

%
%

\markboth{Journal of \LaTeX\ Class Files,~Vol.~14, No.~8, August~2015}%
{Shell \MakeLowercase{\textit{et al.}}: Bare Demo of IEEEtran.cls for IEEE Journals}



\maketitle

\begin{abstract}
Throughput Prediction is one of the primary preconditions for the uninterrupted operation of several network-aware mobile applications, namely video streaming. Recent works have advocated using Machine Learning (ML) and Deep Learning (DL) for cellular network throughput prediction. In contrast, this work has proposed a low computationally complex simple solution which models the future throughput as a multiple linear regression of several present network parameters and present throughput. It then feeds the variance of prediction error and measurement error, which is inherent in any measurement setup but unaccounted for in existing works, to a Kalman filter-based prediction-correction approach to obtain the optimal estimates of the future throughput. Extensive experiments across seven publicly available 5G throughput datasets for different prediction window lengths have shown that the proposed method outperforms the baseline ML and DL algorithms by delivering more accurate results within a shorter timeframe for inferencing and retraining. Furthermore, in comparison to its ML and DL counterparts, the proposed throughput prediction method is also found to deliver higher QoE to both streaming and live video users when used in conjunction with popular Model Predictive Control (MPC) based adaptive bitrate streaming algorithms.
\end{abstract}


%
\IEEEpeerreviewmaketitle

\input{Sections/Introduction}
\input{Sections/Exploratory_Data_Analysis.tex}
\input{Sections/Methodology.tex}
\input{Sections/Results}
\input{Sections/Conclusion.tex}


%




\ifCLASSOPTIONcaptionsoff
  \newpage
\fi



%


\bibliographystyle{IEEEtran}
{\footnotesize
\bibliography{Kalman_main.bib}}

\input{acronyms}
\end{document}

%% file: Sections/Introduction.tex
\section{Introduction}
\IEEEPARstart{T} HE pervasive deployment of ultra high-speed 5G and the migration to 6G \cite{Kato2020,Kato2023} has escalated interest in ultra high bandwidth, ultra low latency applications, such as,  360$^o$ video streaming, volumetric video streaming, live-streaming, \ac{AR}, \ac{VR}, High Definition Map and Image sharing for coordinated driving, etc. Users of such applications demand a high and precise \ac{QoE}, to deliver which the applications tune themselves to the lower layer network throughput. For example, \ac{ABR} \ac{VoD} streaming or live-streaming uses the network throughput to choose the video playback quality such that the user's \ac{QoE} is maximized. 
Such Network Aware Applications (NAAs)~\cite{Ramadan2021,Mondal2020} can, therefore, benefit considerably from the accurate predictions of the future network throughput.\\
\indent Throughput prediction has been extensively investigated ~\cite{Schmid2021} -- for wired networks~\cite{Koutsonikolas2009}, Ethernet~\cite{Jain2002}, WiFi Networks, and cellular networks~\cite{Yue2018, Zhang2019, Raca2019,  Raca2020_1, Narayanan2020, Narayanan2021,Elsherbiny2020}. As network throughput is essentially time series data~\cite{Adhikari2013}, hence, several related prediction methods~\cite{Makridakis2018}, such as, \ac{MA}~\cite{Adhikari2013}, \ac{ARMA}, \ac{ARIMA}~\cite{elsherbiny20204g},  harmonic mean~\cite{Yin2015}, \ac{EWMA}~\cite{Adhikari2013},   etc., have been explored.  However, the throughput of cellular networks depends on a  complex interaction between lower layer  parameters, such as signal strength, frequency of handovers,  \ac{UE} speed, location, neighbouring environment, load of the connected base-station, etc, as well as radio resource allocation algorithms, and upper layer parameters, such as the historical network throughput~\cite{Mondal2020,Elsherbiny2020}. To capture the effect of all these network features on the throughput prediction algorithm, several works have suggested the use of \ac{ML} and \ac{DL} based prediction models~\cite{Yue2018, Raca2019,Raca2020,Minovski2021, Elsherbiny2020,Narayanan2021, Mondal2020,  Schmid2019}.\\
\indent A wide variety of algorithms have been studied in this context. 
 Prediction of \ac{WCDMA} network throughput has been carried out in ~\cite{Nasri2019} using linear regression. The throughput prediction algorithm proposed in \cite{Yue2018} has used \ac{RF} learning with lower layer parameters like \ac{RSSI}, \ac{RSRP}, \ac{CQI} as well as upper layer historical throughput as  input features. The  \ac{RF} based throughput prediction algorithm  in~\cite{Raca2018_2} uses the historical information on throughput and other network parameters to predict the average throughput over a finite future time window. Instead of feeding the entire historical information of the different network parameters,~\cite{Raca2018_2} uses a data summarization technique  which  feeds the mean, inter-quartile range, and the 90th percentile points as input. Several regression based algorithms, like \ac{MLR}, decision tree regression, gradient boosted regression trees, \ac{KNN} regression, \ac{SVR} have been compared in~\cite{Hameed2021} for network-aware \ac{IoT} applications. Of these, \ac{MLR} has been reported to outperform all the other algorithms. The throughput prediction algorithms designed in~\cite{Kousias2019} use \ac{RF}, \ac{MLR}, and \ac{SVR} to strike a balance between prediction accuracy and over-the-air data consumption for mobile broadband networks.  \\
\indent \ac{LSTM}, a variant of recurrent neural networks, has been proposed in~\cite{Schmid2019} for a location independent throughput prediction approach. The authors in~\cite{Raca2020} have also explored throughput prediction  using \ac{LSTM}, along with other \ac{ML} algorithms, viz. \ac{RF} and \ac{SVR}. For the dataset of \cite{Yue2018}, the performance of \ac{LSTM} has been compared with \ac{KNN}, \ac{SVR}, Ridge Regression, \ac{RF}, and \ac{ARIMA} in~\cite{elsherbiny20204g}. Interestingly, \ac{LSTM} outperformed the other algorithms in~\cite{Raca2020}, whereas it is reported in~\cite{elsherbiny20204g} that \ac{RF} delivers the highest accuracy.  The spatio-temporal variability in the network throughput has  been captured in~\cite{Jiantao2020,Chauting2019} using a combination of \ac{LSTM} and \ac{CNN}.  
\\
\indent  Throughput prediction in 5G  has been studied in~\cite{Narayanan2020,Narayanan2021,Minovski2021}. In~\cite{Narayanan2020} is proposed Lumos-5G, a context-aware throughput prediction framework based on  \ac{GBDT} and sequence-to-sequence algorithms. The work  in \cite{Narayanan2021} has evaluated the accuracy of throughput prediction by evaluating its impact on the \ac{QoE} offered by a state-of-art \ac{ABR} video streaming algorithm, i.e., \ac{FMPC}~\cite{Yin2015}. FMPC uses predicted network throughput to decide the future video chunk bitrates.
Both~\cite{Narayanan2020,Narayanan2021} have inferred that \ac{GBDT} outperforms the other algorithms. On the other hand,~\cite{Minovski2021} has inferred that \ac{XGBoost} and \ac{MLP} yield higher accuracies than the traditional \ac{ML} algorithms for their 5G throughput dataset.  Thus, an important takeaway is that although several algorithms have been tried on different datasets, no single model has been found to deliver the highest accuracy of throughput prediction consistently across all datasets. \\
\indent {Notably, throughput prediction methods are not specified by the standards. They are designed to work at the client device and, therefore, may need timely predictions, especially for mobile users.} Most of the existing methods, namely \ac{LSTM}, \ac{CNN}, \ac{MLP}, may fall short of delivering timely estimates on-the-go due to their associated complexity, especially in 5G.  Mathematical modelling, simulations, and measurement-based studies of 5G mmWave networks have revealed that their high frequency of operation makes them limited to line of sight (LoS) communication and renders them susceptible to incessant fluctuations of the underlying link condition~\cite{Narayanan2020_5GOphers}. The time variability of a wireless channel depends on its carrier frequency and is quantified by its coherence time, over which the channel impulse response remains statistically invariant~\cite{Stuber1996}. So, the \ac{DL} models should generate the predicted throughput values within the coherence time of the wireless channel. The 50\% coherence time\footnote{The 50\% coherence time is  $T_c = \sqrt{\frac{9}{16\pi f_m^2}}$. The maximum frequency $f_m$ for velocity $v$, carrier frequency $f_c$, and speed of light $c$ is $f_m=\frac{v}{c}f_c$.} of a vehicular user moving at 36 Km/hr in a 5G network with a carrier frequency of 28 GHz is approximately 0.2 ms. On the other hand, it is reported in~\cite{Raca2020} that the average run time of a \ac{DL} model in an off-the shelf smartphone is 22 ms.  Hence, the channel characteristics may change by the time the \ac{ML} and \ac{DL} based throughput prediction engines deliver inferences. This leaves room for simpler prediction models which will yield timely and accurate results, particularly for 5G and beyond technologies.\\
\indent An inherent assumption in all the existing works is that the measured throughput is accurate.
However, like any other measurement setup, measured throughput is also prone to measurement error or measurement noise~\cite{Belenki2000,Peletta2005}. 
Some of the popular tools used to collect  throughput data include Ookla speedtest\footnote{https://www.speedtest.net/}~\cite{Narayanan2021} and GNetTrackPro\footnote{https://www.gyokovsolutions.com/G-NetTrack\%20Android.html}~\cite{Raca2020_1}. The throughput recorded by these tools depends on the mobile device, the software setups (example, browser or dedicated apps), and the corresponding network connection. The accuracy of the data, therefore, depends largely on the antenna sensitivity of the receivers, and the method adopted by the software tool for recording throughput~\cite{Ookla2022}. Simultaneous connections may also affect the throughput recorded by an application running in a UE~\cite{Ookla2019}. Furthermore, it has also been reported that the ability of GNetTrackPro to record all the metrics is different in different mobile phones and depends on the chipset manufacturer~\cite{Narayanan2021}.   Existing works~\cite{Raca2020,Raca2019,Minovski2021, Elsherbiny2020,Narayanan2021, Mondal2020, Yue2018, Schmid2019} have not accounted for modelling and subsequently cancelling the effect of this measurement noise. In this work, we hypothesize that if the error in measuring the throughput  can be considered  and cancelled, then throughput prediction can be done using considerably simpler models than the complicated \ac{ML} and \ac{DL} models discussed above.\\ 
\indent 
Towards this objective and also in keeping with the Occam's Razor principle~\cite{Occam1996}, in this work, we propose a fresh approach to predict cellular network throughput using a simple multiple linear regression model. However, such a choice will lead to prediction errors or prediction noise~\cite{Hameed2021}. Therefore, we adopt a Kalman filter based prediction-correction approach to obtain the optimal estimate of throughput by cancelling the effect of measurement errors and prediction errors~\cite{Kalman1960}. 
Our proposed model, henceforth referred to as \ac{KFTP}, is simple and computationally less complex than the \ac{ML} and \ac{DL} models. As a result, the training time and the inferencing time of the proposed \ac{KFTP} will also be quite less. This allows the model to be energy-efficient and allows it to be trained at periodic intervals at the end-user device itself, thereby improving the user experience. It also helps the device to deliver timely and accurate results. This will not only benefit ultra reliable and ultra low latency applications of traditional land mobile communications, but also Unmanned Aerial Vehicle (UAV) networks~\cite{Hadiwardoyo2020}. \\
\indent We have tested the accuracy of our \ac{KFTP} using the seven 5G throughput datasets outlined in Table \ref{tab:dataset_det}. For each of these datasets, we have obtained the measurement noise by filtering the data using a moving average filter. This filtered throughput has been assumed to be the true throughput. To justify the use of multiple linear regression, we have undertaken a detailed statistical analysis of the datasets.
  The difference between the true throughput and the throughput predicted by the linear regression model gives the prediction error.  The variances of the measurement and prediction errors have been fed to the Kalman filter to obtain the optimal throughput estimates.\\
\indent Extensive experiments across the datasets, for different filtering window sizes and different prediction windows, show that \ac{KFTP} delivers consistently high $R^2$ scores at par or even better than \ac{ML} or \ac{DL} based algorithms. The algorithm performs particularly well when the environment becomes more noisy, i.e., when the filtering window size is high.
\begin{table*}
 \centering
 \caption{{Details of the Throughput Datasets Used in this work.}}\label{tab:dataset_det}
\begin{tabular}{|c|c|c|c|c|c|}
\hline
\textbf{Dataset} & \textbf{Collected in} & \textbf{Smartphone Models}           & \textbf{Network Service Provider and Type} & \textbf{Mobility} & \textbf{Application}\\ \hline
\textbf{MNWILD-VER}      & Minneapolis           & Samsung Galaxy S20 Ultra 5G  & Verizon, 5G Default                        & Walking         & File Download \\ \hline
\textbf{MNWILD-TNSA }    & Minneapolis           & Samsung Galaxy S20 Ultra 5G  & TMobile, 5G Non-Stand Alone                & Walking     & File Download     \\ \hline
\textbf{MNWILD-TSA}      & Minneapolis           & Samsung Galaxy S20 Ultra 5G  & TMobile, 5G Stand Alone                    & Walking      & File Download    \\ \hline
\textbf{MIWILD-VER}      & Ann Arbor           & Samsung Galaxy S20 Ultra 5G  & Verizon, 5G Default                    & Static      & File Download    \\ \hline
\textbf{LUMOS-5G}         & Minneapolis           & Samsung Galaxy S10                   & Verizon, 5G Non-Stand Alone                & Driving      & Video Streaming   \\ \hline
\textbf{IRISH-DD}         & Unspecified Irish City       & Samsung Galaxy S10                   & An Irish Mobile Network Provider, 5G       & Driving      & File Download   \\ \hline
\textbf{IRISH-DS}         & Unspecified Irish city       & Samsung Galaxy S10                   & An Irish Mobile Network Provider, 5G       & Static      & File Download     \\ \hline
\end{tabular}
\end{table*}
To analyze the applicability of \ac{KFTP}, we have used it as the throughput prediction engine in -1) \ac{ABR} \ac{VoD} streaming and in 2) live streaming. For \ac{VoD} streaming, we have compared the \ac{QoE} performance of the combination of \ac{FMPC}~\cite{Yin2015} and the proposed \ac{KFTP} against a combination of \ac{FMPC} with other baseline throughput prediction algorithms viz. \ac{ARIMA}, \ac{SVR}, \ac{RF}, \ac{XGBoost}, and \ac{LSTM}.  \ac{KFTP} has been observed to outperform others, especially by reducing the rebuffering time. Similarly, we have compared the \ac{QoE} offered by the live video streaming algorithm Live-MPC~\cite{monkeysun2019}, when it is used with our proposed \ac{KFTP} algorithm as opposed to the other baseline algorithms. We have seen that \ac{KFTP} delivers the highest \ac{QoE} by reducing the bitrate fluctuation and the latency.  \\
\indent  \textit{Paper Organization: }\sectionlabel\ref{sec:Exploration} analyses the interrelation between throughput data and various network parameters. \sectionlabel\ref{sec:Methodology} explains the proposed \ac{KFTP}. \sectionlabel\ref{sec:Results} discusses the results on the accuracy of \ac{KFTP}. \sectionlabel\ref{sec:VS} discusses the impact of \ac{KFTP} in improving the \ac{QoE} of \ac{VoD} and live streaming in 5G. \sectionlabel\ref{sec:Conclusion} concludes the paper.

\

%% file: Sections/Exploratory_Data_Analysis.tex
\section{Exploratory Data Analysis} \label{sec:Exploration}

\indent In this section, we describe various benchmarking datasets containing 5G throughput data. Our target is to design a generalized throughput prediction algorithm to forecast future throughput values across different environments. Hence, we have undertaken an extensive study with diverse datasets collected from different locations, using different handsets that are connected to different service providers. Through this study we have explored several statistical properties which establish relationships of different 5G network parameters with throughput. Such relationships may be integral towards designing the target throughput prediction algorithm. In the following, we provide a brief description of the popular datasets that have been extensively used in the present study.
\subsubsection{\textbf{MN-Wild}}
The dataset in \cite{Narayanan2021} contains 5G throughput collected in Minneapolis, MN, where  service providers Verizon and TMobile have both deployed 5G networks. The detailed description of MN-Wild dataset is provided in \cite{Narayanan2021}. We have further divided the MN-Wild dataset into three subsets, as  in Table \ref{tab:dataset_det}. 
\subsubsection{\textbf{MI-Wild}}
This dataset contains 5G throughput data collected from Ann Arbor, Michigan, for stationary \acp{UE}~\cite{Narayanan2021}. 
 
 \subsubsection{\textbf{Lumos 5G}}
 Lumos 5G~\cite{Narayanan2020} contains data collected from the Verizon 5G Non-Stand Alone (NSA) network using a Samsung Galaxy S10 mobile phone, from different parts of Minneapolis. The data belongs to a video streaming application under various mobility conditions (static, walking, driving), collected at a traffic intersection, inside an airport, and along a driving loop. A subset of the Lumos-5G dataset is available online\footnote{https://lumos5g.umn.edu/}, which has been used in this work.
\subsubsection{\textbf{Irish}}
The \textit{Irish} dataset in~\cite{Raca2020_1} contains 5G Throughput from an Irish mobile network provider collected using Samsung Galaxy S10 while walking and driving, considering file download and video streaming workloads. We have divided it into two subsets as in Table \ref{tab:dataset_det}. \\
\indent These datasets provide real time measurements of network features of 5G, collected through  experiments and have been extensively used in literature for throughput prediction in 5G ~\cite{Narayanan2021,Narayanan2020,Narayanan2020_5GOphers}. However, any set of values recorded through experiments is associated with measurement noise. As discussed earlier, the error or noise in measurement is primarily contributed by the measurement setup, such as the antenna sensitivity of the smartphones, the chipsets used in the phones, and the application setups. So, in this work, we pre-process the throughput data to remove the measurement noise.\\

\begin{table}[h]
\centering
\caption{{Correlation Coefficient ($\rho$) between Measured and Filtered Throughput of the Datasets in Table \ref{tab:dataset_det}, for Different  Filter Window sizes, $F$, and prediction window length \predlength{5}.} }
\label{tab:filtered_raw_data_corr_doeff}

\begin{tabular}{|l|ccc|}
\hline
\multicolumn{1}{|c|}{\multirow{2}{*}{Dataset}} & \multicolumn{3}{c|}{\begin{tabular}[c]{@{}c@{}}Filter Window Size (F) \\ in seconds\end{tabular}} \\ \cline{2-4} 
\multicolumn{1}{|c|}{}                         & \multicolumn{1}{c|}{\textbf{3}}        & \multicolumn{1}{c|}{\textbf{5}}       & \textbf{7}       \\ \hline
\textbf{MNWILD-VER}                            & \multicolumn{1}{c|}{0.956}             & \multicolumn{1}{c|}{0.934}            & 0.917            \\ \hline
\textbf{MNWILD-TNSA}                           & \multicolumn{1}{c|}{0.935}             & \multicolumn{1}{c|}{0.895}            & 0.868            \\ \hline
\textbf{MNWILD-TSA}                            & \multicolumn{1}{c|}{0.95}              & \multicolumn{1}{c|}{0.919}            & 0.898            \\ \hline
\textbf{MIWILD-VER}                            & \multicolumn{1}{c|}{0.961}             & \multicolumn{1}{c|}{0.928}            & 0.898            \\ \hline
\textbf{LUMOS-5G}                              & \multicolumn{1}{c|}{0.949}             & \multicolumn{1}{c|}{0.906}            & 0.869            \\ \hline
\textbf{IRISH-DD}                              & \multicolumn{1}{c|}{0.906}             & \multicolumn{1}{c|}{0.789}            & 0.702            \\ \hline
\textbf{IRISH-DS}                              & \multicolumn{1}{c|}{0.897}             & \multicolumn{1}{c|}{0.774}            & 0.696            \\ \hline
\end{tabular}

\end{table}

\begin{figure}[t]
    \centering
\includegraphics[width=0.4\textwidth]{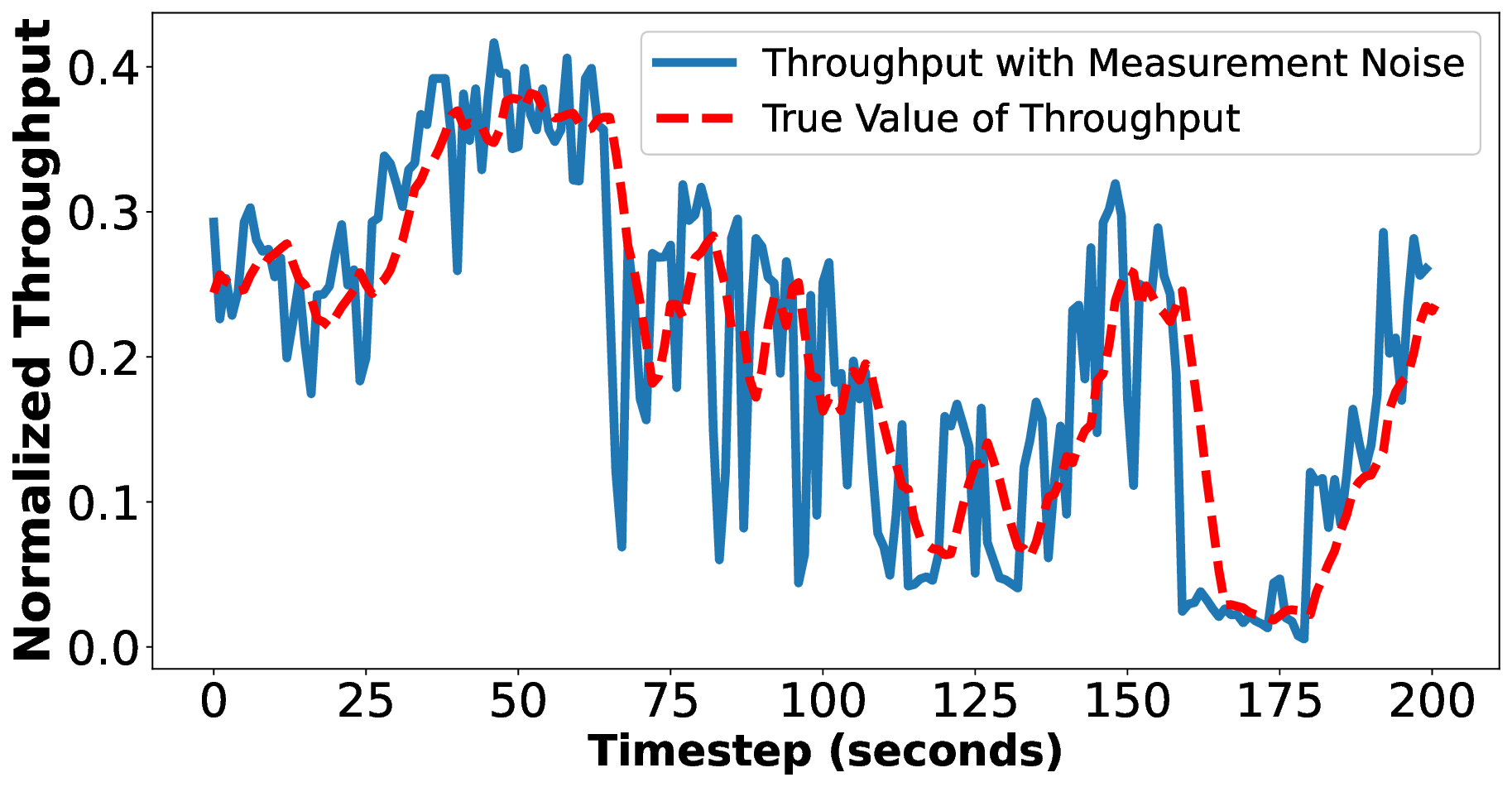}
    \caption{{Measured and True throughput  of the MNWILD-VER Dataset for \filterwindow{7} samples.}}
    \label{fig:measured_vs_true_timeseries}
\end{figure}

\begin{figure}[t] 
    \centering\includegraphics[width=0.3\textwidth]{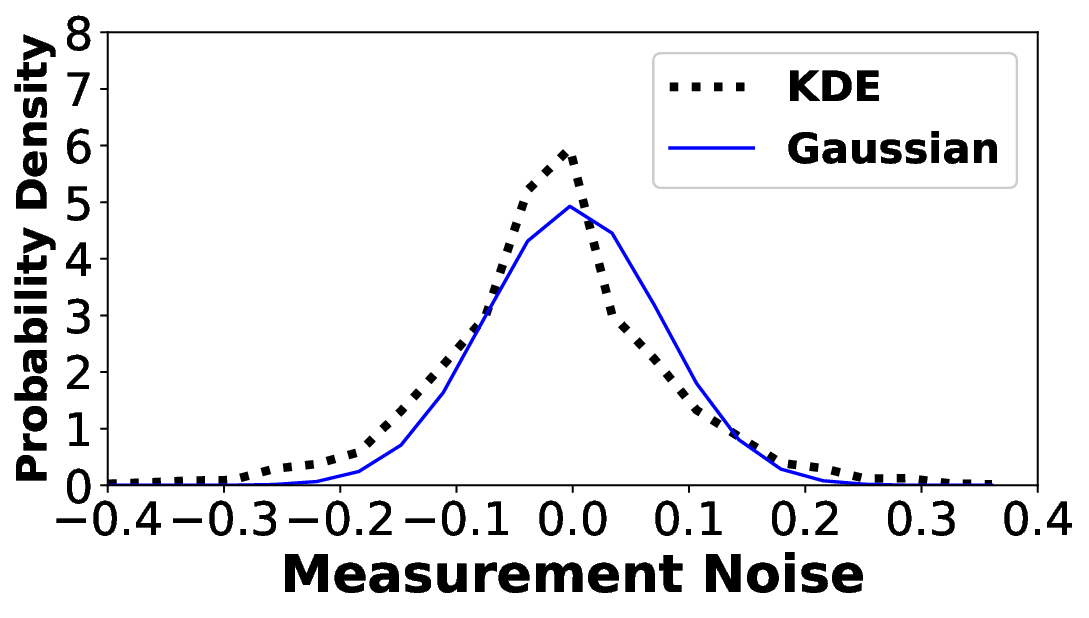}
    \caption{{KDE of the Measurement Error in the MNWILD-VER dataset for \filterwindow{7} samples is zero mean Gaussian with standard deviation $\sigma_M=0.0632$.}}
    \label{fig:ME_kde_vs_gauss}
\end{figure}

\indent We have adopted a reasonable assumption of filtering the time-series data using a moving average filter of a very small window size ($F$). It may be observed from Table \ref{tab:filtered_raw_data_corr_doeff} that the correlation coefficient between the measured data and the throughput data for a filter window  of \filterwindow{3} samples is almost equal to 0.9 for all seven datasets. Additionally, the correlation coefficient is greater than 0.85 for \filterwindow{5,7} for all the datasets except IRISH-DD and IRISH-DS. The high correlation coefficient values between the measured throughput and the filtered throughput indicates that no significant loss of information has occurred in the preprocessing step. Therefore, our assumption of obtaining the true value of throughput data does not involve distorting the original trend of the measured time series. Rather a noise signal of very small power, which we assume to be the measurement noise, is eliminated in this filtering step. Fig. \ref{fig:measured_vs_true_timeseries} shows the time series of the measurement (raw) throughput and the filtered throughput of the MNWILD-VER dataset for \filterwindow{7} samples. It is seen that the measured and filtered throughput have similar trends and characteristics. Thus, in this work, we have considered the resultant filtered data to be the true value of throughput. \\
\indent The window size ($F$) is, therefore, an important parameter of our proposed algorithm. A detailed parametric analysis on throughput prediction for varying sizes of $F$ is presented in \sectionlabel\ref{sec:Results}. The present approach has two benefits - one, it takes into account the noise of the measurement device, and two, it prevents such measurement noise from affecting the prediction algorithm adversely. \\
\indent We have assumed that the measurement noise is additive and stationary. The \ac{PDF} of the normalized measurement noise has been estimated using  a rectangular kernel. The \ac{KDE} of this noise is found to closely follow a zero mean Gaussian random variable. For example, the KDE of the measurement noise corresponding to Fig. \ref{fig:measured_vs_true_timeseries} and a Gaussian fit with mean$ = 0$ and standard deviation $\sigma_M=0.06$ is shown in Fig. \ref{fig:ME_kde_vs_gauss}. The obtained mean square error is in the order of $10^{-2}$.\\
\indent Our objective in this work is to predict the throughput $L$ seconds into the future. So, after pre-processing, we have performed a statistical analysis of the filtered data to identify the potential network parameters which influence the throughput. For this, we have evaluated linear correlation coefficient ($\rho$) of the present network parameters and present throughput with the future throughput values for a time difference of $L$ seconds. 
Table \ref{tab:pearsoncoeff} shows the value of $\rho$ between the present network parameters, such as speed, RSRP, RSRQ, SINR, throughput, and the future throughput of all the seven datasets, for a time difference of \predlength{5} seconds in the future.
\begin{table}[t]
\centering
\caption{ {Pearson's Correlation  Coefficient `$\rho$' of the present network parameters with future throughput for \predlength{5} secs, for the datasets of Table I. Unit of future throughput is in bits/sec. Units of  other features are mentioned in the table.}}
\begin{tabular}{|p{2.1cm}|lllll|}
\hline

 & \multicolumn{1}{p{0.7cm}|}{\textbf{Present Speed (m/s)}} & \multicolumn{1}{p{0.7cm}|}{\textbf{Present RSRP (dB)}} & \multicolumn{1}{p{0.7cm}|}{\textbf{Present RSRQ (dB)}} & \multicolumn{1}{p{0.7cm}|}{\textbf{Present SINR (dB)}} & \multicolumn{1}{p{1.2cm}|}{\textbf{Present Throughput (bps)}} \\ \hline
\textbf{MNWILD-VER} & \multicolumn{1}{l|}{-0.05} & \multicolumn{1}{l|}{0.22} & \multicolumn{1}{l|}{N/A} & \multicolumn{1}{l|}{0.72} & 0.93 \\ \hline
\textbf{MNWILD-TNSA} & \multicolumn{1}{l|}{0.11} & \multicolumn{1}{l|}{0.23} & \multicolumn{1}{l|}{N/A} & \multicolumn{1}{l|}{-0.18} & 0.87 \\ \hline
\textbf{MNWILD-TSA} & \multicolumn{1}{l|}{-0.12} & \multicolumn{1}{l|}{0.58} & \multicolumn{1}{l|}{N/A} & \multicolumn{1}{l|}{-0.02} & 0.90 \\ \hline
\textbf{MIWILD-VER} & \multicolumn{1}{l|}{N/A} & \multicolumn{1}{l|}{0.42} & \multicolumn{1}{l|}{N/A} & \multicolumn{1}{l|}{0.38} & 0.87 \\ \hline
\textbf{LUMOS-5G} & \multicolumn{1}{l|}{-0.25} & \multicolumn{1}{l|}{0.49} & \multicolumn{1}{l|}{-0.18} & \multicolumn{1}{l|}{0.50} & 0.84 \\ \hline
\textbf{IRISH-DD} & \multicolumn{1}{l|}{-0.25} & \multicolumn{1}{l|}{0.29} & \multicolumn{1}{l|}{-0.02} & \multicolumn{1}{l|}{0.12} & 0.63 \\ \hline
\textbf{IRISH-DS} & \multicolumn{1}{l|}{N/A} & \multicolumn{1}{l|}{-0.15} & \multicolumn{1}{l|}{0.06} & \multicolumn{1}{l|}{-0.05} & 0.66 \\ \hline

\end{tabular}
\label{tab:pearsoncoeff}
\end{table}
\begin{figure}[t]
     \centering
     \includegraphics[width=0.3\textwidth]{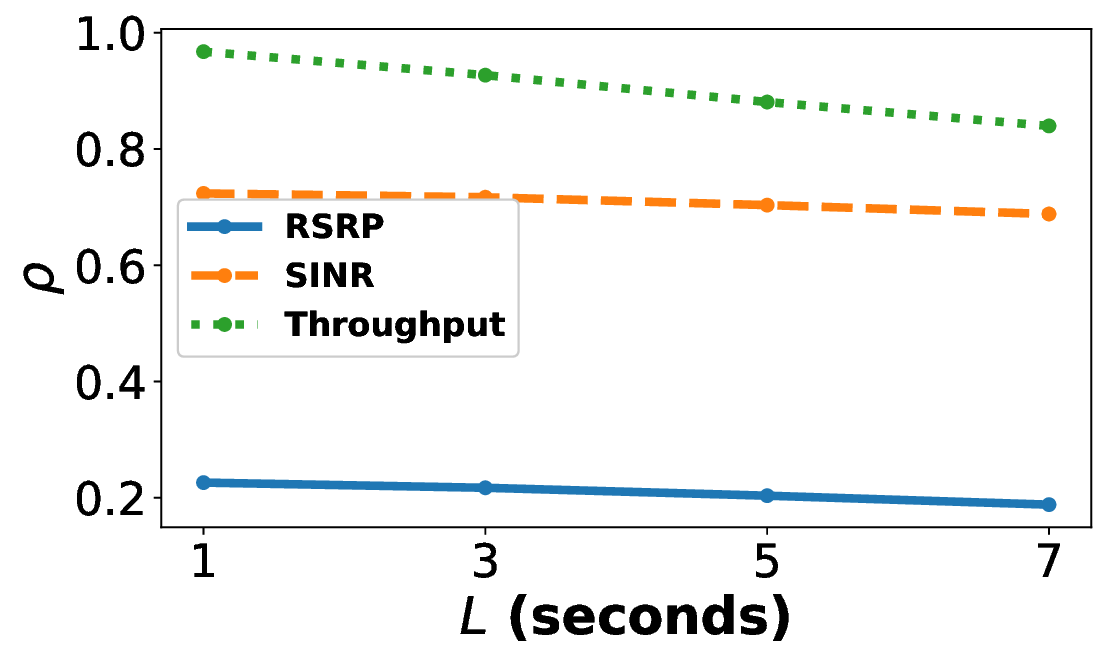}
    \caption{{Correlation Coefficient ($\rho$) vs. time lead (L) for different network parameters of MNWILD-VER dataset.}}
    \label{fig:pearson-vs-lag-plot}
\end{figure}
\begin{figure*}[t]
     \centering
     \hspace{-0.2cm}
    \begin{subfigure}{0.25\textwidth}
        \centering
        \includegraphics[width = \textwidth]{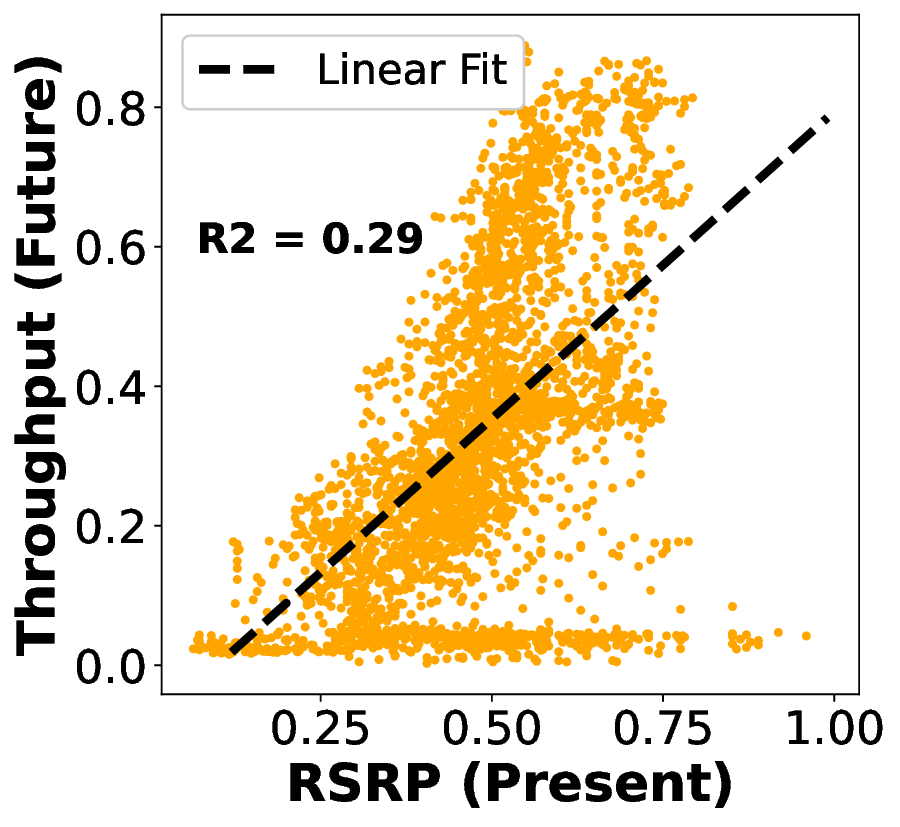} 
        \caption{{Normalized RSRP.}}
        \label{fig:RSRP-scatter}
    \end{subfigure}
     \hspace*{0.2cm}
     \begin{subfigure}{0.25\textwidth}
         \centering
         \includegraphics[width = \textwidth]{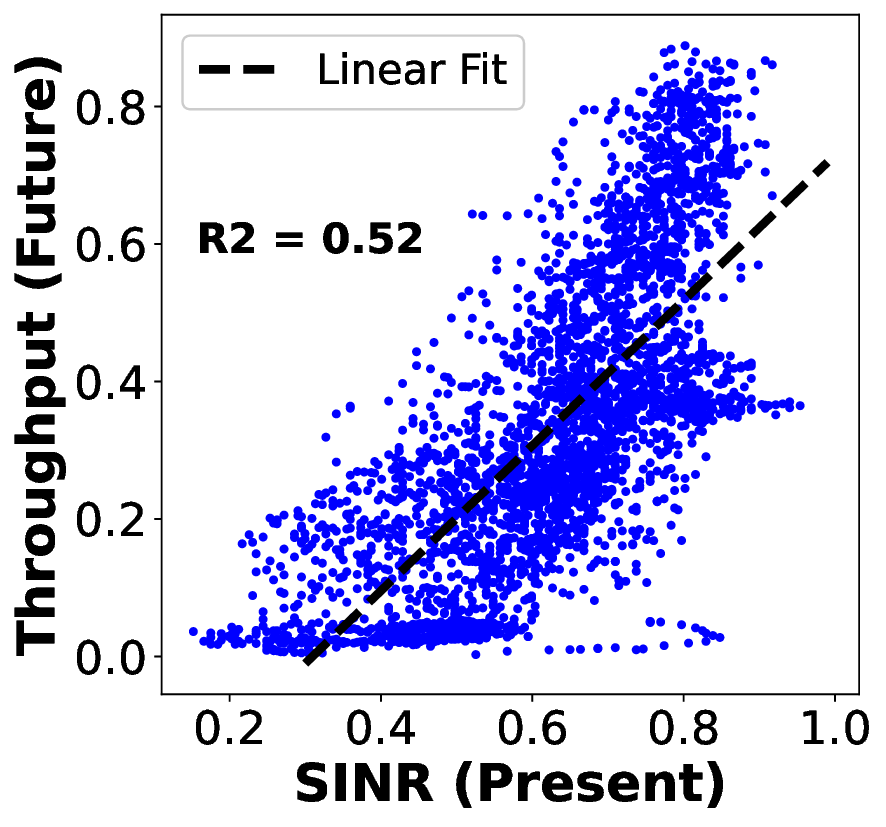}
        \caption{{Normalized SINR.}}
        \label{fig:SNR-scatter}
     \end{subfigure}
     \hspace*{0.2cm}
     \begin{subfigure}{0.25\textwidth}
         \centering
         \includegraphics[width = \textwidth]{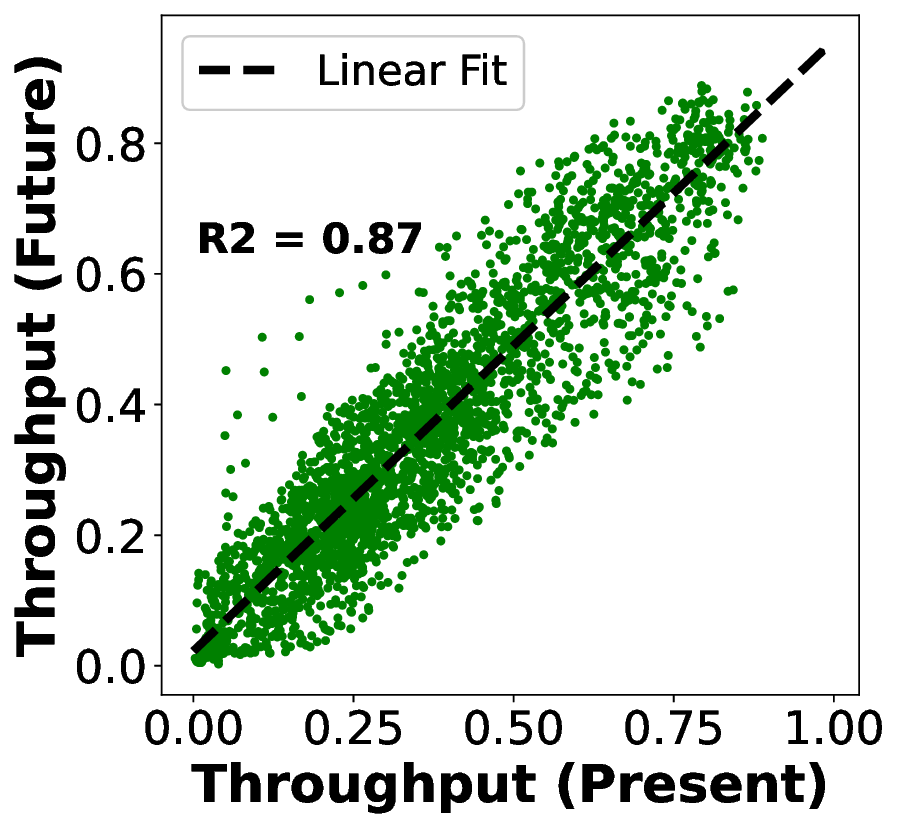}
        \caption{{Normalized Throughput.}}
        \label{fig:Throughput-scatter}
     \end{subfigure}   \centering\caption{{Future Normalized Throughput vs. Present Network Features; Dataset -  MNWILD-VER, \predlength{5} secs, \filterwindow{7} samples.}}
\label{fig:thpt_vs_params_scatter}
\end{figure*}
It may be noted that the correlation of the future throughput is the highest with the present  network throughput. In some datasets, the correlation of the future throughput with the present \ac{RSRP} and \ac{SINR} is also quite high $(\rho>0.5)$. Data on other network parameters, such as \ac{RSRQ} and \ac{CQI}, are, however, not consistently available across all datasets, and in those in which these features are present, the reported $\rho$ is not significantly high. The speed of movement, even though present for all seven datasets, never shows a substantially high correlation coefficient with the future throughput values. Therefore, for designing our throughput prediction algorithm under different locations and different network conditions, we have considered three key network parameters, viz. present throughput value, \ac{RSRP}, and \ac{SINR}. It is also evident from  Table \ref{tab:pearsoncoeff} that value of $\rho$ between any present network parameter and future throughput values are different for different datasets, i.e., for different network conditions. Furthermore, the value of $\rho$ also depends on the time difference $L$. As an example, in Fig. \ref{fig:pearson-vs-lag-plot}, we have shown the effect of $L$ on the value of $\rho$  for the MNWILD-VER dataset.\\
\indent To establish a relationship between the present RSRP, SINR, and present throughput with the future throughput, we have shown their scatter plots in Fig. \ref{fig:thpt_vs_params_scatter}, for the MNWILD-VER dataset with a time lead of \predlength{5} seconds. Observing the nature of these plots and the correlation coefficient values, we have adopted the assumption of a linearity also shown in Fig. \ref{fig:thpt_vs_params_scatter}. 
The $R^2$ scores for linear fitting of the present RSRP, present \ac{SNR}, and present throughput with the future throughput are 0.29, 0.52, and 0.87, respectively. These $R^2$ scores indicate that the variation in future throughput is substantially captured by the variation of the chosen present network features. Thus, we propose a prediction-correction based throughput prediction algorithm using linear state equations which we discuss in the following section. 

%% file: Sections/Methodology.tex
\section{Methodology} \label{sec:Methodology}
In this section, we present a new Kalman Filter based Throughput Prediction (KFTP) algorithm for predicting 5G  throughput from real-time  data collected at mobile \acp{UE}. As explained before, any real-time measurement is associated with measurement noise. \ac{KFTP} is, therefore, designed using a prediction and correction approach following the traditional Kalman filter paradigm~\cite{Kalman1960}, to cancel the effect of measurement noise. Although it has been observed from literature that complicated prediction models provide accuracy in a particular environment, they have failed to perform consistently across several datasets. In the present work, we have used a simple Multiple Linear Regression (\ac{MLR})  for the prediction step. This not only saves our algorithm from the problem of over fitting, but also reduces the overall time complexity associated with training the model. The predicted value obtained from the linear state equation is further updated in the correction step to get the optimal throughput by compensating for the prediction error as well as the measurement error.  \\
\indent Let $x(n)$ denote the true throughput at the 
time-step $n$. Let the network parameters \ac{RSRP} and \ac{SINR} at the time-step $n$ be denoted as $u_1(n)$ and $u_2(n)$, respectively. Our objective is to predict $x(n+L)$ from the values of $u_1(n)$, $u_2(n)$ and $x(n)$. Hence, the feature vector at the time-step $n$ is represented as  $\textbf{y}(n+L) = [1 \ u_1(n) \  u_2(n) \ x(n)]^T$. Therefore, considering the linearity assumption of \sectionlabel\ref{sec:Exploration}, the predicted throughput at time-step $n+L$ is $\hat{x}(n+L)=\mathbf{a}^T\textbf{y}$, i.e., 
\begin{equation} \label{eq:LSE}
\hat{x}(n+L)=a_0+a_1 u_1(n)+a_2 u_2(n)+ a_3 x(n)
\end{equation}
\begin{figure}
    \centering \includegraphics[width=0.3\textwidth]{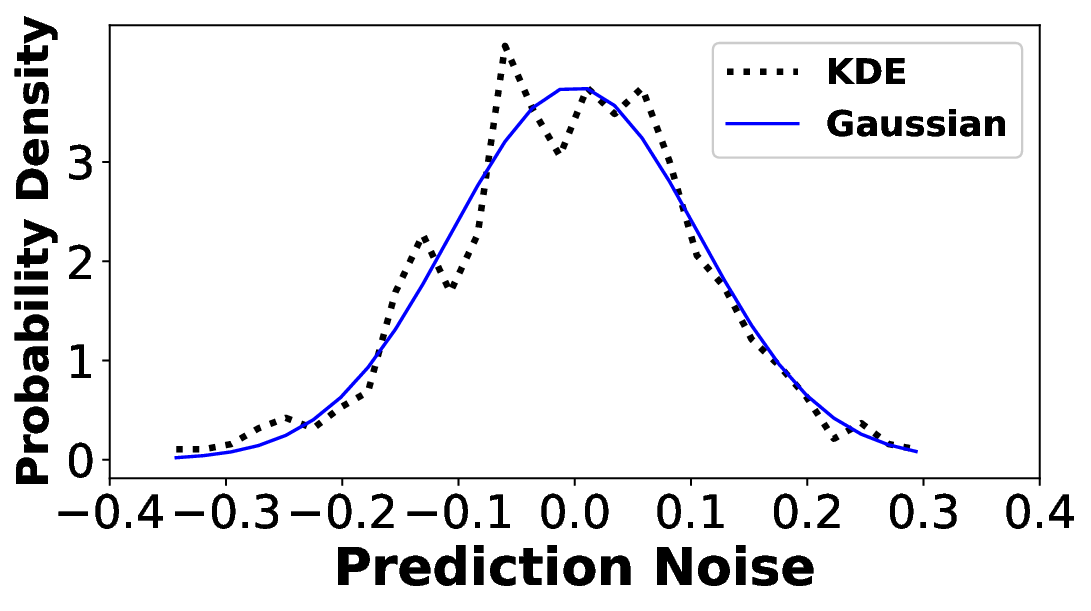}
    \caption{{KDE of the Prediction Error in the MNWILD-VER dataset for \predlength{5} seconds and  \filterwindow{7} samples is zero mean Gaussian with standard deviation $\sigma_P=0.08$.}}
    \label{fig:PE_kde_vs_gauss}
\end{figure}
\indent To find the optimized values of the coefficients, $\mathbf{a}= [a_0$ $a_1$ $a_2$ $a_3]^T$, the least squares estimation approach has been adopted such that the \ac{RSS} is minimized. i.e.,  
\begin{align}\label{eq:RSS}
    \underset{\mathbf{a}}{min}\ \mathbf{RSS}(\mathbf{a}) = \left(x(n) - \sum_{i} a_i y_i\right)^2. 
\end{align}
\indent (\ref{eq:LSE}) represents the prediction step of our model. Once the coefficients are optimized for a given dataset, they remain unchanged thereafter. As each dataset corresponds to one smartphone connected to one service provider in a given city engaged in one specific mobile application, the coefficients remain unchanged for one such setup. It will be different for different smartphone models, different service providers, different cities, and different mobile phone applications. \\
\begin{table}[h]
\caption{{Symbol Notations.}}\label{tab:symbol-key}
\begin{tabular}{|p{1.5cm}|p{6.5cm}|}
\hline
\textbf{Symbols}        & \textbf{Definition}                                      \\ \hline
$\tilde{u}(i,n)$ $\forall i\in\{1,2\}$ & Measured value of: 1) \ac{RSRP} and 2) \ac{SINR} at time step $n$\\ \hline
$\tilde{x}(n)$         & Measured value of throughput at time step $n$            \\ \hline
$\hat{x}(n)$            & Predicted value of throughput  at time step $n$          \\ \hline
$P(n)$                 & Predicted Variance of  throughput at time step $n$       \\ \hline
$\hat{x}^{*}(n)$       & Corrected/Estimated value of throughput at time step $n$ \\ \hline
$P^{*}(n)$               & Estimated variance of throughput at time step $n$        \\ \hline
$\sigma^2_P,\sigma^2_M$ & Variance of Prediction Noise/Measurement Noise           \\ \hline
$S(n)$                        & Covariance of Measured and predicted throughput at time-step $n$           \\ \hline
$K_n$                        & Kalman gain at time-step $n$                              \\ \hline
\end{tabular}
\end{table}

\indent The difference between the output of the \ac{MLR} and true throughput gives the   \textit{prediction error}. We show that the prediction error is well characterized using a Gaussian PDF. Fig. \ref{fig:PE_kde_vs_gauss} shows the KDE plot of the prediction error  and its  corresponding Gaussian approximation. The mean square error for the Gaussian fit (zero mean and standard deviation $\sigma_P=0.0836$) has been found to be in the order of $10^{-2}$. Once the prediction step, and the \ac{PDF} of the measurement and the prediction errors are obtained, the optimal throughput $\hat{x^*}(n)$ is obtained using the Kalman filtering methodology~\cite{Kalman1960}. 
\begin{algorithm}[t]
\SetAlgoLined
\caption{{Kalman Filter based Throughput Prediction (KFTP).}}\label{algo: KF}
\textbf{Input:} Prediction noise variance $\sigma^2_P$, Measurement noise variance $\sigma_M^2$, time lead $L$, the coefficient vector $\mathbf{a}$ from the linear regression in (\ref{eq:LSE})\;
\For{$n \in \{1,2,3,...,N\}$}
{
    \If{($n\leq L$)} 
    {
    $\hat{x}^*(n),\hat{x}(n) \gets \tilde{x}(n)$; {/*Initialize using measured throughput*/}\texttt{\\}\label{algol:estthuptoL}
    $P^*(n),P(n),K(n)\gets 0$\;\label{algol:covmeaspredthuptoL}
    $S(n) \gets \sigma^2_M$\;}
    \If{($n\geq L+1$)}
    {
      $S(n) \gets P(n) + \sigma^2_M$; {/*Calculate Covariance*/}\texttt{\\} \label{algol:covmeaspred}
    $K(n) \gets P(n)/S(n)$; {/*Calculate Kalman Gain*/}\texttt{\\}\label{algol:calc_KG}
    $\hat{x}^{*}(n) \gets \hat{x}(n) + K(n)\times(\tilde{x}(n)-\hat{x}(n)$); 
 {/*Obtain Optimal Estimate*/}\texttt{\\}\label{algol:optthest}
    $P^{*}(n)\gets (1-K(n))\times P(n)$; {/*Obtain Variance*/}\texttt{\\}\label{algol:optVar}
    }
    $\hat{x}(n+L) \gets a_3\hat{x}^*(n)+[a_0 \ a_1 \ a_2]\mathbf{u(n)}$; {/*Predict Throughput*/}\texttt{\\}\label{algol:prth}
    $P(n+L) \gets a^2_3 P^{*}(n)+\sigma^2_P$\;\label{algol:prvar}
}
\end{algorithm}
The steps of our proposed KFTP algorithm  are  outlined in Algorithm \ref{algo: KF}. The corresponding symbol notations are provided in Table \ref{tab:symbol-key}. The inputs to our algorithm are -- 
\begin{enumerate*}
    \item present values of network throughput $\tilde{x}(n)$,
    \item present values of network parameters, viz. \ac{RSRP}, and \ac{SINR},$\mathbf{u(n)}$,
    \item the measurement noise variance, $\sigma^2_M$,
    \item the prediction noise variance, $\sigma^2_P$,
    \item length of the prediction window or time lead, $L$,
    \item the coefficient vector, $\mathbf{a}$, obtained by solving (\ref{eq:RSS}).
\end{enumerate*}

At time-step $n$, the future throughput $\hat{x}(n+L)$ is predicted  using the optimal throughput estimate, $x^*(n)$. Then, the predicted throughput $\hat{x}(n+L)$ is corrected using the measured throughput $\tilde{x}(n+L)$ to obtain the optimal estimate  $\hat{x}^*(n+L)$. As $\hat{x}(n+L)$ is predicted using $\hat{x}^*(n)$ and not by $\tilde{x}(n)$ or $\hat{x}(n)$, the effects of both the measurement error due to the measurement setup and the modelling error are minimized. {In this context it is important to note that our KFTP algorithm involves four sets of computation as indicated in Line 11 (one multiplication and two addition), Line 12 (one multiplication, one addition), Line 14 (four multiplication, three addition) and Line 15 (two multiplication, one addition) of Algorithm 1. Thus, the present KFTP algorithm requires only eight multiplicative and seven additive computations to produce an optimal prediction of future throughput.}

%% file: Sections/Results.tex
\section{Results and Discussion}\label{sec:Results}
\begin{figure*}
    \centering
    \begin{subfigure}{0.2\textwidth}
    \includegraphics[width=1\textwidth]{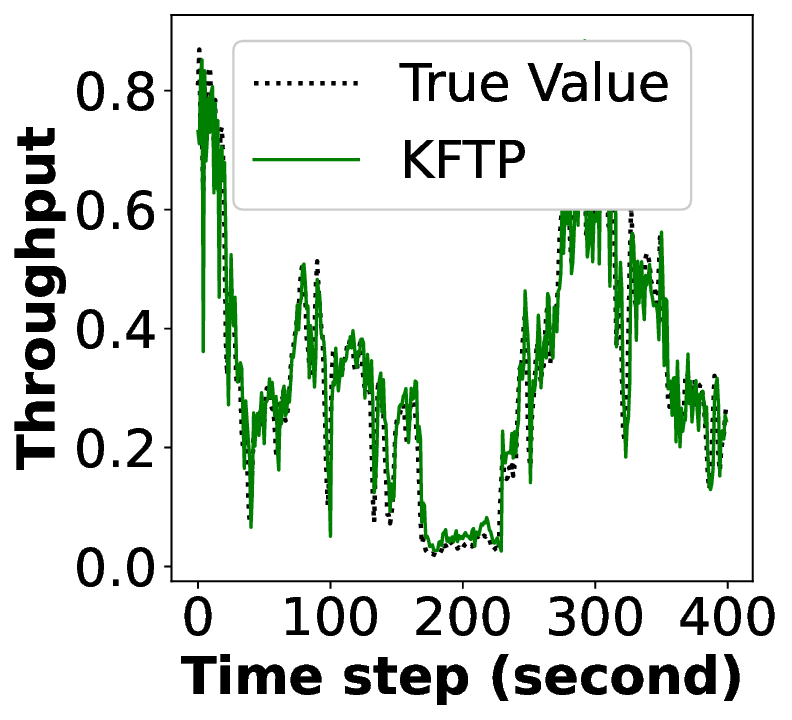}
            \caption{{\textbf{MNWILD-VER.}}}
            \label{fig:KALMAN_TIMESERIES_VER}
    \end{subfigure}
    \begin{subfigure}{0.2\textwidth}
            \includegraphics[width=1\textwidth]{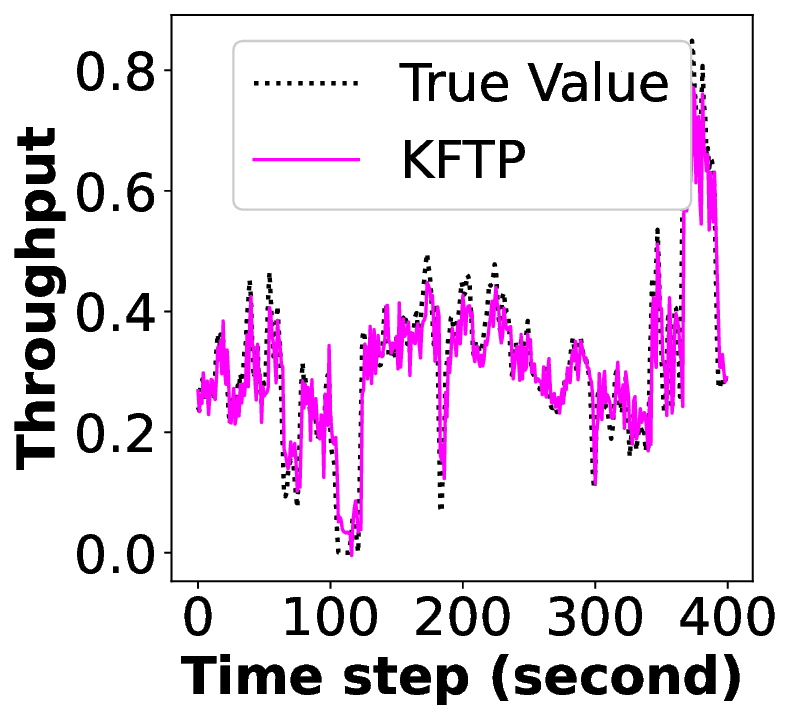}
            \caption{{\textbf{MNWILD-TNSA.}}}
            \label{fig:KALMAN_TIMESERIES_TNSA}
    \end{subfigure}
    \begin{subfigure}{0.2\textwidth}
            \includegraphics[width=1\textwidth]{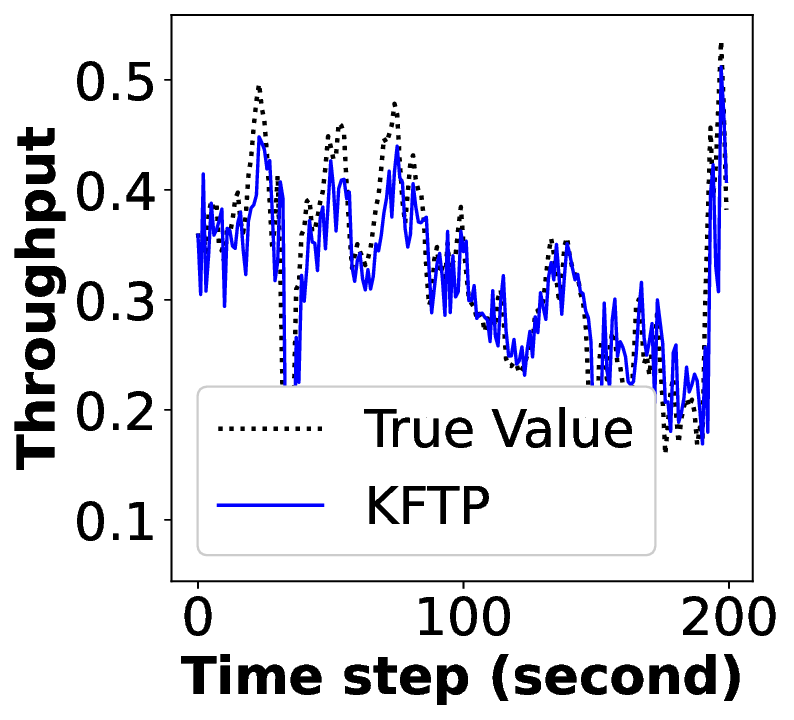}
            \caption{{\textbf{MNWILD-TSA.}}}
            \label{fig:KALMAN_TIMESERIES_TSA}
    \end{subfigure}
    \begin{subfigure}{0.2\textwidth}
            \includegraphics[width=1\textwidth]{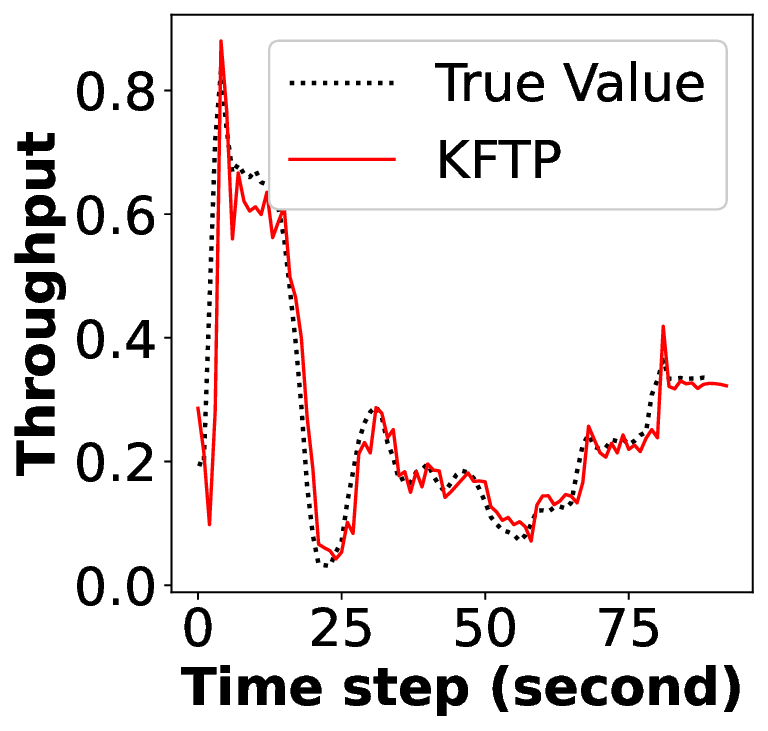}
            \caption{{\textbf{MIWILD-VER.}}}
            \label{fig:KALMAN_TIMESERIES_MIWILDVER}
    \end{subfigure}
    
    \begin{subfigure}{0.2\textwidth} 
            \includegraphics[width=1\textwidth]{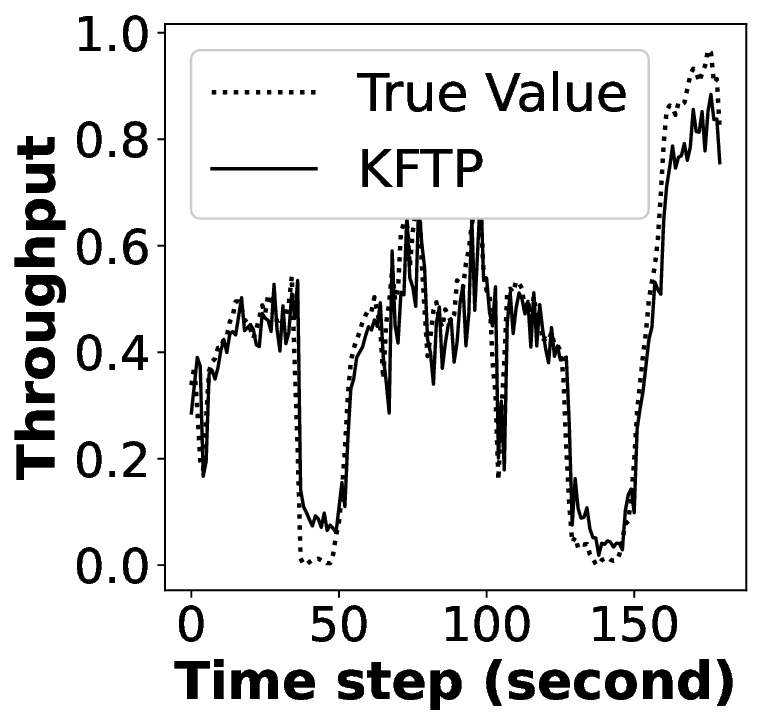}
            \caption{\textbf{{LUMOS-5G.}}}
            \label{fig:KALMAN_TIMESERIES_LUMOS5G}
    \end{subfigure}
    \begin{subfigure}{0.2\textwidth} 
            \includegraphics[width=1\textwidth]{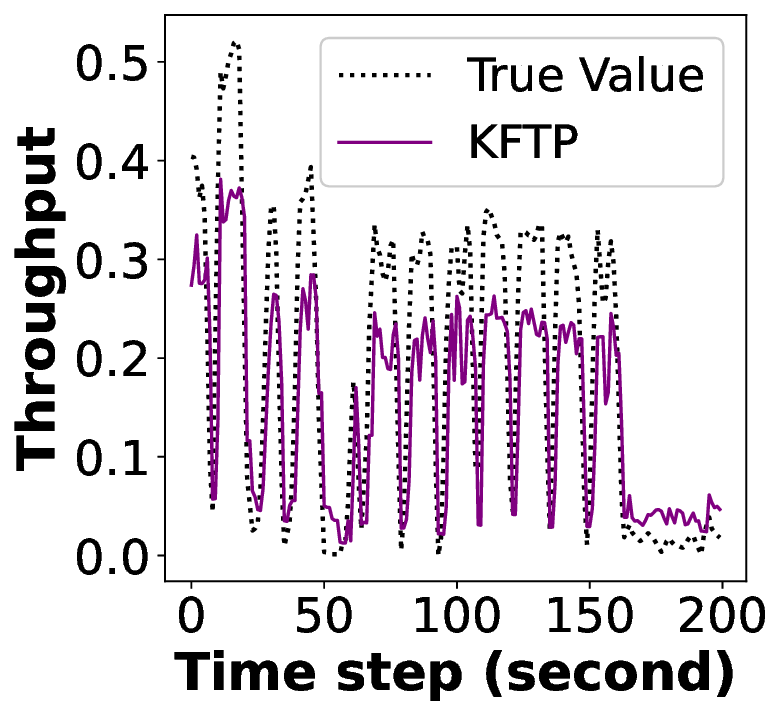}
            \caption{{\textbf{IRISH-DD.}}}
            \label{fig:KALMAN_TIMESERIES_IRISH_DD}
    \end{subfigure}
    \begin{subfigure}{0.2\textwidth} 
            \includegraphics[width=1\textwidth]{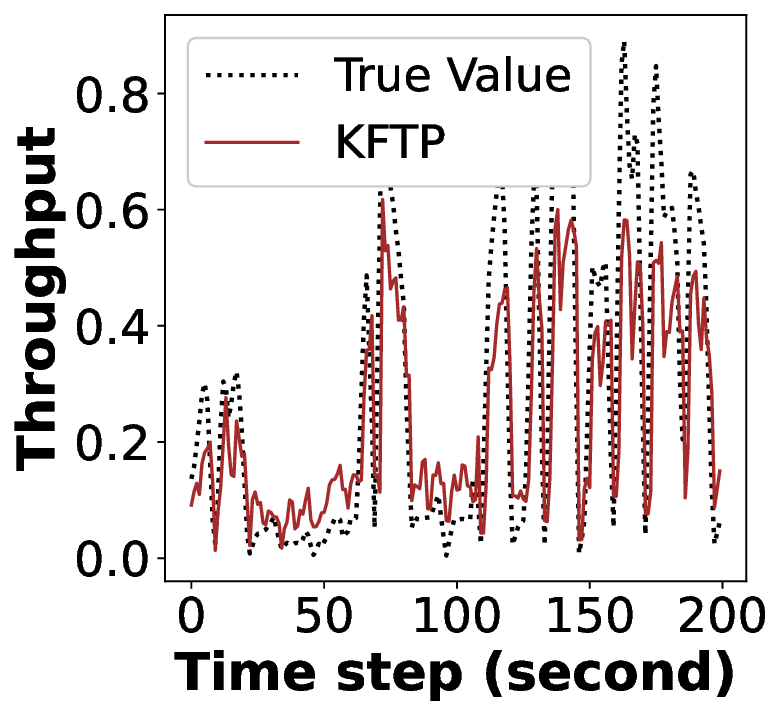}
            \caption{{\textbf{IRISH-DS.}}}
            \label{fig:KALMAN_TIMESERIES_IRISH_DS}
    \end{subfigure}

    \caption{{Time series plot of Normalized Throughput - KFTP vs. Optimal value (\predlength{3} secs, \filterwindow{3} samples).}}
    \label{fig:KALMAN_TIMESERIES}
\end{figure*}
In this section, we have first presented the simulation setup of the throughput prediction algorithms. We have then established the efficacy of \ac{KFTP} by comparing it with other baseline algorithms.
\subsection{Simulation Setup}
We have used the publicly available 5G datasets tabulated in Table \ref{tab:dataset_det} to train and test our \ac{KFTP}. The train-test split of the datasets has been kept at 80\%- 20\%. This implies that the linear regression coefficients $\mathbf{a}$ have been obtained by training it with 80\% of all the data points. The remaining  20\% sample points have been used as the testing set on which the performance of the proposed \ac{KFTP} has been evaluated.  
\begin{table*}
\footnotesize
\centering
\caption{\label{tab:R2_COMPARISON_TRUE} {Comparison of Proposed \ac{KFTP} with Baseline Throughput Prediction algorithms for \filterwindow{3} samples.}}
\begin{tabular}{|c|c|cc|cc|cc|cc|cc|} 
\hline
\multirow{3}{*}{\textbf{Datasets}}     & \multirow{3}{*}{\begin{tabular}[c]{@{}c@{}}\textbf{Throughput}\\ \textbf{Prediction}\\ \textbf{Algorithm}\end{tabular}} & \multicolumn{10}{c|}{\textbf{Time Lead, L (seconds)}}                                                                                                                              \\ 
\cline{3-12}
                             &                                                                                              & \multicolumn{2}{c|}{\textbf{1}}         & \multicolumn{2}{c|}{\textbf{3}}         & \multicolumn{2}{c|}{\textbf{5}}         & \multicolumn{2}{c|}{\textbf{7}}         & \multicolumn{2}{c|}{\textbf{9}}          \\ 
\cline{3-12}
                             &                                                                                              & $\textbf{R}^\textbf{2}$          & \textbf{MAE}           & $\textbf{R}^\textbf{2}$          & \textbf{MAE}           & $\textbf{R}^\textbf{2}$          & \textbf{MAE}           & $\textbf{R}^\textbf{2}$          & \textbf{MAE}           & $\textbf{R}^\textbf{2}$         & \textbf{MAE}            \\ 
\hline
\multirow{6}{*}{\textbf{MNWILD-VER}}  & KFTP                                                                                         & \textbf{0.89} & \textbf{0.05} & \textbf{0.89} & \textbf{0.05} & \textbf{0.88} & \textbf{0.06} & 0.87          & \textbf{0.06} & \textbf{0.86} & \textbf{0.06}  \\
                             & ARIMA~\cite{raca2017back}                                                                                        & 0.88          & \textbf{0.05} & 0.86          & 0.06          & 0.83          & \textbf{0.06} & 0.81          & \textbf{0.06} & 0.75          & 0.07           \\
                             & SVR~\cite{Raca2020}                                                                                          & 0.88          & 0.06          & 0.88          & 0.06          & 0.87          & \textbf{0.06} & 0.86          & \textbf{0.06} & 0.84          & 0.07           \\
                             & RF~\cite{Yue2018}                                                                                           & 0.87          & 0.06          & 0.84          & 0.07          & 0.82          & 0.07          & 0.81          & 0.08          & 0.75          & 0.08           \\
                             & XGBoost~\cite{Minovski2021}                                                                                     & 0.86          & 0.06          & 0.84          & 0.07          & 0.82          & 0.07          & 0.8           & 0.07          & 0.75          & 0.08           \\
                             & LSTM\cite{Mei2020}                                                                                          & 0.88          & 0.06          & 0.88          & 0.06          & 0.87          & \textbf{0.06} & \textbf{0.88} & \textbf{0.06} & 0.83          & 0.07           \\ 
\hline
\multirow{6}{*}{\textbf{MNWILD-TNSA}} & KFTP                                                                                         & \textbf{0.82} & \textbf{0.04} & \textbf{0.82} & \textbf{0.04} & \textbf{0.8}  & \textbf{0.04} & \textbf{0.77} & \textbf{0.04} & \textbf{0.74} & \textbf{0.04}  \\
                             & ARIMA~\cite{raca2017back}                                                                                         & \textbf{0.82} & \textbf{0.04} & 0.79          & \textbf{0.04} & 0.76          & \textbf{0.04} & 0.71          & \textbf{0.04} & 0.66          & 0.05           \\
                             & SVR~\cite{Raca2020}                                                                                          & 0.81          & \textbf{0.04} & 0.79          & \textbf{0.04} & 0.73          & \textbf{0.04} & 0.7           & 0.05          & 0.61          & 0.05           \\
                             & RF~\cite{Yue2018}                                                                                           & \textbf{0.82} & \textbf{0.04} & \textbf{0.82} & \textbf{0.04} & \textbf{0.8}  & 0.05          & \textbf{0.77} & 0.05          & \textbf{0.74} & 0.06           \\
                             & XGBoost~\cite{Minovski2021}                                                                                      & \textbf{0.82} & \textbf{0.04} & 0.78          & \textbf{0.04} & 0.72          & \textbf{0.04} & 0.67          & 0.05          & 0.61          & 0.05           \\
                             & LSTM~\cite{Mei2020}                                                                                          & 0.81          & \textbf{0.04} & 0.81          & \textbf{0.04} & \textbf{0.8}  & \textbf{0.04} & 0.76          & 0.04          & 0.71          & 0.05           \\ 
\hline
\multirow{6}{*}{\textbf{MNWILD-TSA}}  & KFTP                                                                                         & 0.87          & \textbf{0.04} & \textbf{0.87} & \textbf{0.04} & \textbf{0.86} & \textbf{0.04} & 0.85          & \textbf{0.04} & 0.83          & 0.05           \\
                             & ARIMA~\cite{raca2017back}                                                                                         & \textbf{0.88} & \textbf{0.04} & 0.84          & \textbf{0.04} & 0.81          & \textbf{0.04} & 0.79          & \textbf{0.04} & 0.75          & 0.05           \\
                             & SVR~\cite{Raca2020}                                                                                          & 0.85          & 0.05          & 0.8           & 0.05          & 0.77          & 0.05          & 0.76          & 0.05          & 0.72          & 0.06           \\
                             & RF~\cite{Yue2018}                                                                                            & 0.85          & \textbf{0.04} & 0.78          & 0.05          & 0.69          & 0.06          & 0.64          & 0.07          & 0.58          & 0.07           \\
                             & XGBoost~\cite{Minovski2021}                                                                                      & 0.85          & \textbf{0.04} & 0.81          & 0.05          & 0.77          & 0.05          & 0.73          & 0.06          & 0.69          & 0.06           \\
                             & LSTM~\cite{Mei2020}                                                                                          & 0.86          & \textbf{0.04} & \textbf{0.87} & \textbf{0.04} & \textbf{0.86} & \textbf{0.04} & \textbf{0.86} & \textbf{0.04} & \textbf{0.86} & \textbf{0.04}  \\ 
\hline
\multirow{6}{*}{\textbf{MIWILD-VER}}  & KFTP                                                                                         & \textbf{0.84} & \textbf{0.04} & \textbf{0.84} & \textbf{0.04} & \textbf{0.84} & 0.05          & 0.83          & \textbf{0.05} & \textbf{0.81} & 0.06           \\
                             & ARIMA~\cite{raca2017back}                                                                                         & \textbf{0.84} & \textbf{0.04} & 0.83          & \textbf{0.04} & 0.8           & \textbf{0.04} & 0.75          & \textbf{0.05} & 0.68          & \textbf{0.05}  \\
                             & SVR~\cite{Raca2020}                                                                                          & 0.83          & 0.05          & 0.8           & 0.06          & 0.77          & 0.06          & 0.69          & 0.07          & 0.63          & 0.08           \\
                             & RF~\cite{Yue2018}                                                                                           & 0.82          & 0.05          & 0.79          & 0.05          & 0.68          & 0.06          & 0.64          & 0.08          & 0.55          & 0.09           \\
                             & XGBoost~\cite{Minovski2021}                                                                                      & 0.76          & 0.06          & 0.62          & 0.08          & 0.41          & 0.09          & 0.27          & 0.1           & 0.2           & 0.12           \\
                             & LSTM~\cite{Mei2020}                                                                                          & 0.83          & \textbf{0.04} & \textbf{0.84} & \textbf{0.04} & \textbf{0.84} & \textbf{0.04} & \textbf{0.84} & \textbf{0.05} & 0.8           & \textbf{0.05}  \\ 
\hline
\multirow{6}{*}{\textbf{LUMOS-5G}}     & KFTP                                                                                         & \textbf{0.93} & 0.05          & \textbf{0.91} & \textbf{0.06} & 0.86          & \textbf{0.08} & 0.79          & 0.11          & 0.71          & 0.12           \\
                             & ARIMA~\cite{raca2017back}                                                                                         & \textbf{0.93} & \textbf{0.04} & 0.9           & \textbf{0.06} & \textbf{0.89} & \textbf{0.08} & \textbf{0.87} & \textbf{0.09} & \textbf{0.84} & \textbf{0.1}   \\
                             & SVR~\cite{Raca2020}                                                                                          & 0.89          & 0.06          & 0.87          & 0.07          & 0.84          & \textbf{0.08} & 0.8           & \textbf{0.09} & 0.74          & \textbf{0.1}   \\
                             & RF~\cite{Yue2018}                                                                                           & 0.9           & 0.05          & 0.82          & 0.07          & 0.73          & 0.09          & 0.63          & 0.11          & 0.53          & 0.13           \\
                             & XGBoost~\cite{Minovski2021}                                                                                      & 0.89          & 0.06          & 0.86          & 0.07          & 0.83          & 0.09          & 0.79          & 0.1           & 0.74          & 0.12           \\
                             & LSTM~\cite{Mei2020}                                                                                          & 0.91          & \textbf{0.04} & 0.9           & \textbf{0.06} & 0.83          & \textbf{0.08} & 0.76          & \textbf{0.09} & 0.69          & \textbf{0.1}   \\ 
\hline
\multirow{6}{*}{\textbf{IRISH-DD}}    & KFTP                                                                                         & 0.75          & 0.06          & 0.77          & 0.07          & 0.61          & 0.1           & 0.42          & 0.13          & 0.5           & 0.12           \\
                             & ARIMA~\cite{raca2017back}                                                                                         & \textbf{0.79} & \textbf{0.05} & \textbf{0.78} & \textbf{0.06} & \textbf{0.78} & \textbf{0.07} & \textbf{0.75} & \textbf{0.08} & \textbf{0.72} & 0.14           \\
                             & SVR~\cite{Raca2020}                                                                                          & 0.76          & 0.07          & 0.55          & 0.1           & 0.14          & 0.13          & 0.07          & 0.13          & 0.09          & 0.13           \\
                             & RF~\cite{Yue2018}                                                                                           & 0.78          & \textbf{0.05} & 0.52          & 0.09          & 0.27          & 0.11          & 0.08          & 0.12          & 0.17          & 0.12           \\
                             & XGBoost~\cite{Minovski2021}                                                                                      & \textbf{0.79} & \textbf{0.05} & 0.69          & 0.08          & 0.33          & 0.11          & 0.13          & 0.12          & 0.2           & 0.12           \\
                             & LSTM~\cite{Mei2020}                                                                                          & \textbf{0.79} & \textbf{0.05} & 0.71          & 0.08          & 0.42          & 0.09          & 0.21          & 0.12          & 0.2           & \textbf{0.11}  \\ 
\hline
\multirow{6}{*}{\textbf{IRISH-DS}}    & KFTP                                                                                         & 0.72          & 0.09          & \textbf{0.74} & 0.11          & 0.71          & 0.14          & 0.68          & 0.16          & 0.6           & 0.15           \\
                             & ARIMA~\cite{raca2017back}                                                                                         & \textbf{0.75} & \textbf{0.08} & \textbf{0.74} & \textbf{0.09} & \textbf{0.74} & \textbf{0.09} & \textbf{0.74} & \textbf{0.09} & \textbf{0.71} & \textbf{0.13}  \\
                             & SVR~\cite{Raca2020}                                                                                          & 0.74          & 0.09          & 0.72          & 0.11          & 0.49          & 0.17          & 0.5           & 0.19          & 0.49          & 0.17           \\
                             & RF~\cite{Yue2018}                                                                                           & 0.73          & 0.09          & 0.54          & 0.13          & 0.26          & 0.17          & 0.13          & 0.19          & 0.26          & 0.17           \\
                             & XGBoost~\cite{Minovski2021}                                                                                      & 0.74          & \textbf{0.08} & 0.69          & 0.13          & 0.59          & 0.17          & 0.47          & 0.18          & 0.55          & 0.17           \\
                             & LSTM~\cite{Mei2020}                                        & 0.71          & \textbf{0.08} & 0.63          & 0.12          & 0.45          & 0.15          & 0.2           & 0.17          & 0.27          & 0.16 \\\hline         
\end{tabular}
\end{table*}
\normalsize
\subsection{Results}
\begin{figure}[t]
    \centering
    \begin{subfigure}{0.23\textwidth}
            \includegraphics[width=1\textwidth]{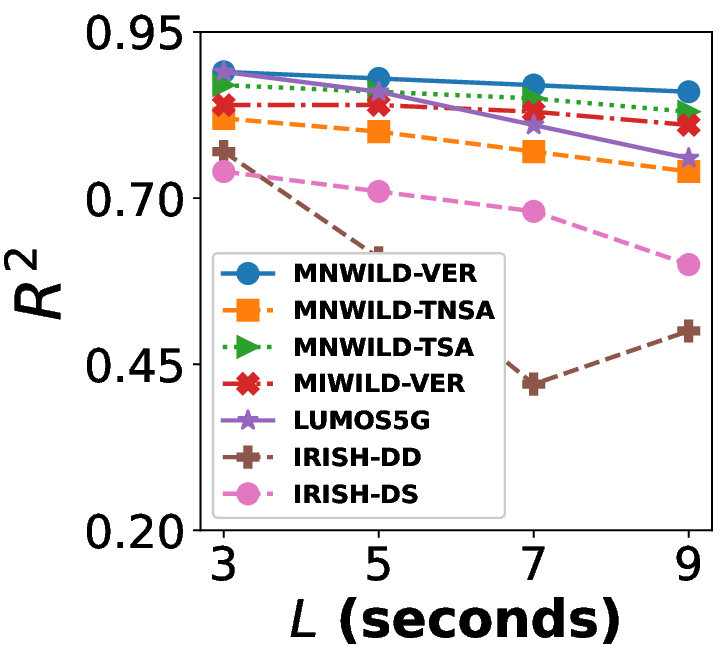}
            \caption{{$R^2$ score.}}
            \label{fig:r2-kftp-all-datasets}
    \end{subfigure}
    \begin{subfigure}{0.23\textwidth}
            \includegraphics[width=1\textwidth]{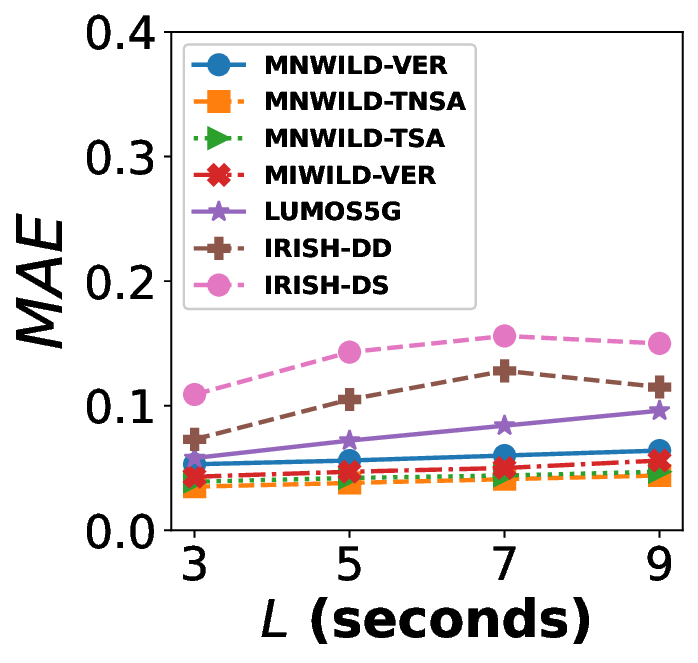}
            \caption{{MAE}}
            \label{fig:mae-kftp-all-datasets}
    \end{subfigure}
    \caption{{Performance of KFTP (\filterwindow{3} samples).}}
    \label{fig:r2-mae-kftp-all-datasets}
\end{figure}

\begin{figure}
    \centering
    \begin{subfigure}{0.23\textwidth}
            \includegraphics[width=1\textwidth]{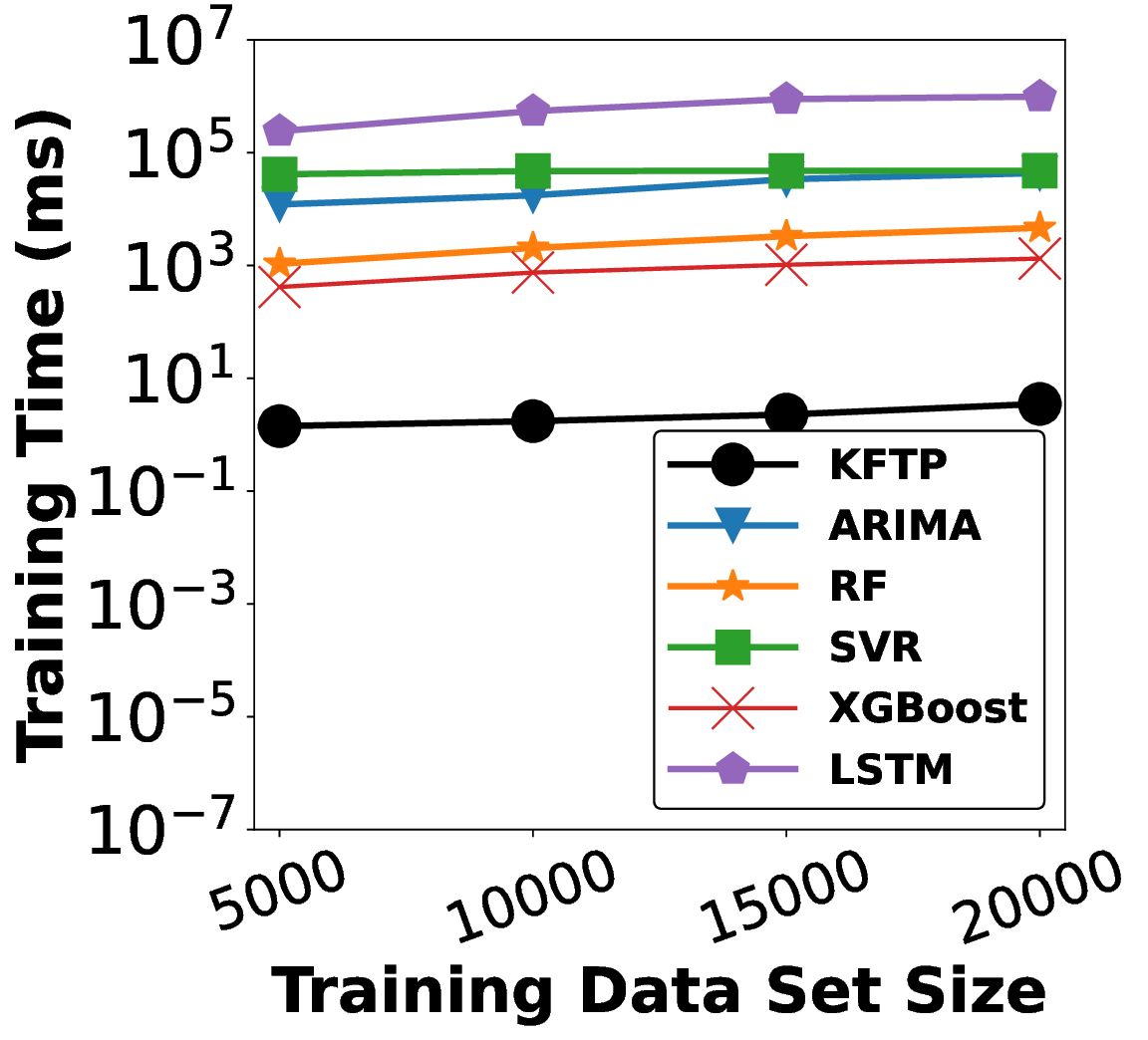}
            \caption{ {Training Time.}}
            \label{fig:complexity-train}
    \end{subfigure}
    \begin{subfigure}{0.23\textwidth}
            \centering            \includegraphics[width=1\textwidth]{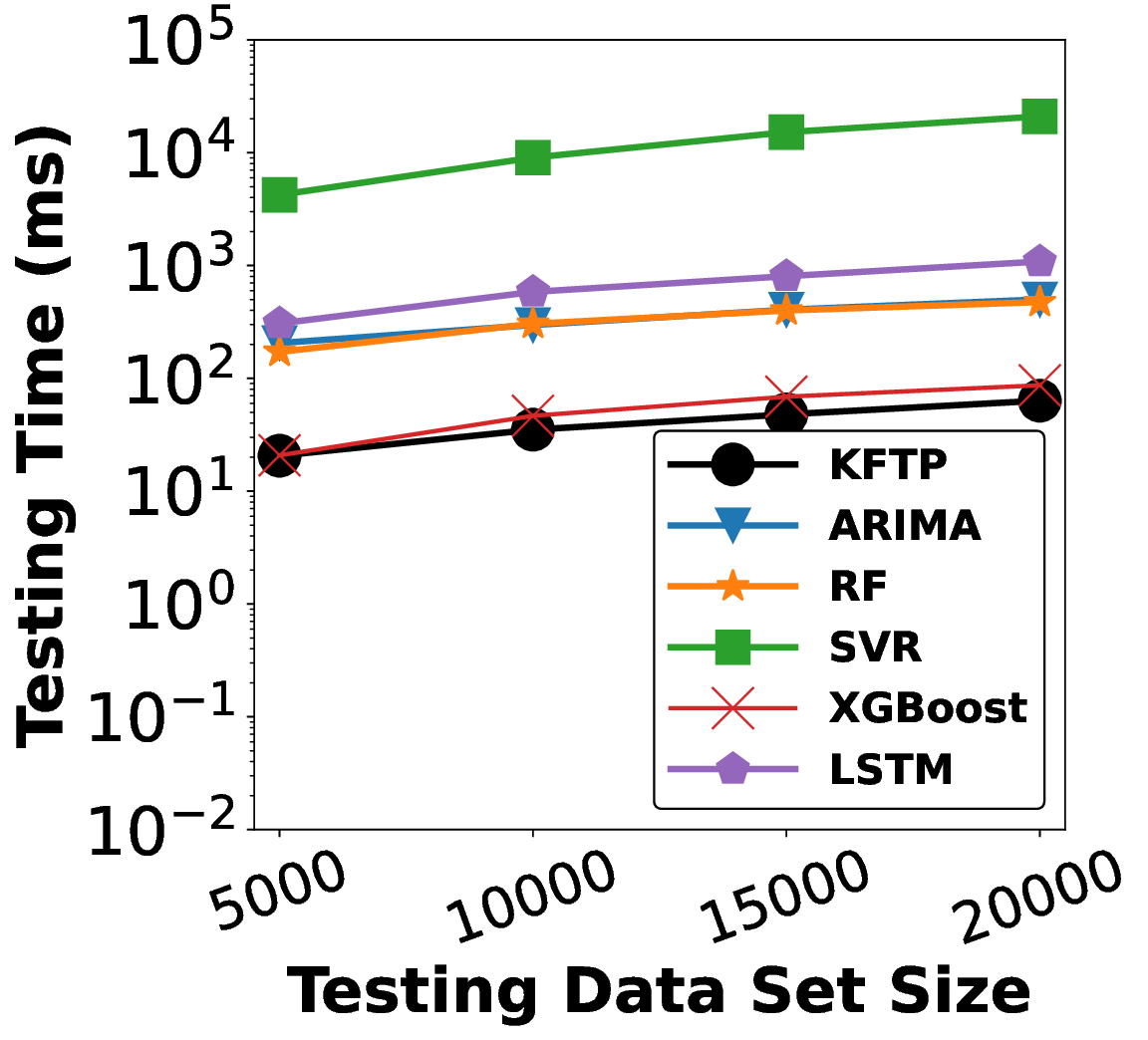}
            \caption{{Testing Time.}}
            \label{fig:complexity-test}
    \end{subfigure}
    \caption{{Computational Time of Throughput Prediction Algorithms for Varying Input Size for F = 3, \predlength{5}. (Dataset: LUMOS-5G); System Specifications: Processor 4 Core, 8 Thread Intel i5, with CPU of 1.6 GHz and 8 GB RAM .}}
    \label{fig:complexity}
\end{figure}
\indent Fig. \ref{fig:KALMAN_TIMESERIES} presents the true throughput values and the optimal estimates of the future throughput generated by \ac{KFTP} for a time lead of \filterwindow{3} seconds corresponding to the datasets in Table \ref{tab:dataset_det}. It is observed that the proposed \ac{KFTP} algorithm performs reasonably for all seven datasets. Two performance evaluation metrics viz. \acf{MAE} and coefficient of regression ($R^2$) have been considered, as outlined in Table \ref{tab:R2_COMPARISON_TRUE}. It may be seen that the \ac{MAE} is always less than 10\% for a time lead of one second. Furthermore, for the same time lead, the $R^2$ score is more than 0.8, except for the IRISH-DD and IRISH-DS datasets. The poor performance for the IRISH datasets can be attributed to the lower correlation coefficients between the present network parameters and the present throughput with the future throughput values, as may be observed from Table~\ref{tab:pearsoncoeff}. In many real-time applications, for example in video streaming, it may be necessary to predict throughput over longer prediction windows~\cite{Mondal2020}. Hence, to demonstrate the robustness and reliability of the \ac{KFTP} algorithm, the $R^2$ scores and \acp{MAE} for  \predlength{3,5,7,9} seconds have also been obtained and are tabulated in Table \ref{tab:R2_COMPARISON_TRUE}. 

The effect of $L$ on $R^2$ score and \ac{MAE} are shown in Fig. \ref{fig:r2-kftp-all-datasets} and \ref{fig:mae-kftp-all-datasets}, respectively. It may be observed that the performance of the \ac{KFTP} algorithm deteriorates with higher values of $L$. This is because the decrease in correlation between the future throughput and the present network features with higher values of $L$ (Fig. \ref{fig:pearson-vs-lag-plot}) compromises the prediction accuracy of our linear state equation in (\ref{eq:LSE}). Therefore, it may be inferred that the  \ac{KFTP} algorithm will not always be able to guarantee a reliable prediction of future throughput for a significantly high $L$ ($L >= 20$s). However, throughput prediction for a horizon of ten seconds can be considered to be sufficient for most practical engineering applications. It may be noted that \ac{KFTP} predicts future throughput with acceptable accuracy within this time range, as may be observed from Figs. \ref{fig:r2-kftp-all-datasets} and \ref{fig:mae-kftp-all-datasets}. \\
\indent To establish the efficacy of \ac{KFTP}, it has been compared with the following five baseline throughput prediction algorithms, across all the datasets in the present study -- 
\begin{enumerate*}
    \item \acf{ARIMA} \cite{raca2017back}, 
    \item \acf{SVR} \cite{Raca2020}, 
    \item \acf{RF} \cite{Yue2018}, \item \acf{XGBoost} \cite{Minovski2021}, \item \acf{LSTM} \cite{Mei2020} 
\end{enumerate*}. The corresponding performance metrics ($R^2$, MAE) are provided in Table \ref{tab:R2_COMPARISON_TRUE}. It is observed that our \ac{KFTP} algorithm performs comparably with other algorithms across all the datasets. Particularly, \ac{KFTP} outperforms the other algorithms for the MNWILD-VER, MIWILD-VER and MNWILD-TNSA datasets. For the other datasets, \ac{KFTP} predicts the future throughput with acceptable \ac{MAE} and $R^2$ score. An important point to note is that Network Aware Applications (NAA) may need the throughput prediction engine to be retrained every time the \ac{UE} visits a new city with a different network scenario. The target throughput prediction algorithm should have smaller training times, especially for the energy-constrained handheld \ac{UE} devices. Additionally, any real-time prediction for applications like live-video streaming will necessitate accurate and timely inferencing. Thus, an efficient throughput prediction algorithm will need to have short training and testing times. Fig. \ref{fig:complexity} shows the training and the testing times taken by all the algorithms discussed so far, for the Lumos-5G dataset. The time lead considered is \predlength{3}, and filter window is \filterwindow{3}. It may be observed from Fig. \ref{fig:complexity} and Table \ref{tab:R2_COMPARISON_TRUE} that the proposed \ac{KFTP} produces reasonably accurate throughput estimates within significantly short training and testing times. This may be attributed to the lower complexity of \ac{MLR} and the fewer computations needed by \ac{KFTP} as outlined in \sectionlabel\ref{sec:Methodology}.\\
\begin{figure}[t]
    \centering
    \begin{subfigure}{0.23\textwidth}
        \includegraphics[width=\textwidth]{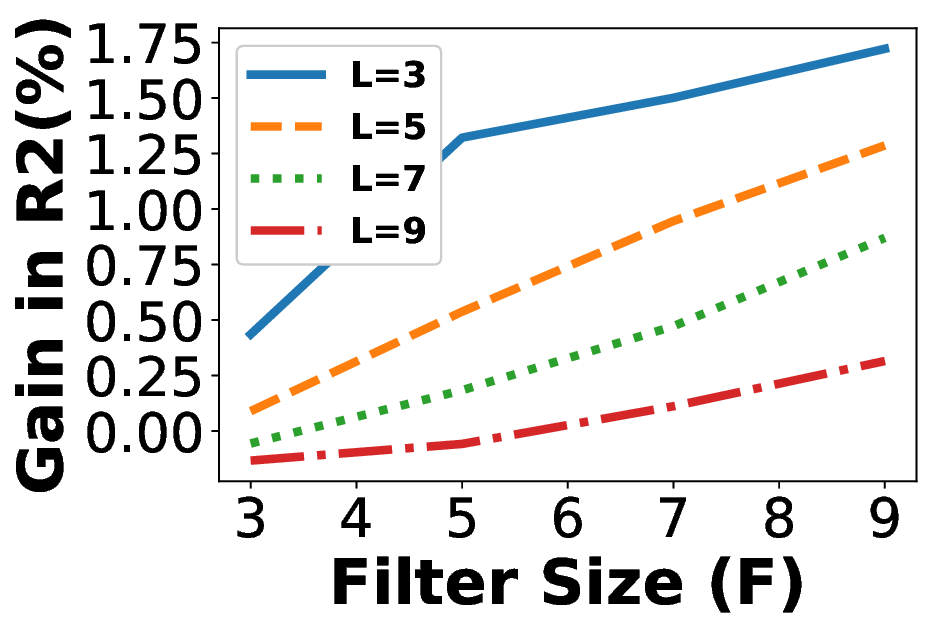}
        \caption{{Dataset: MNWILD-VER.}}
        \label{fig:r2vsfilter_ver}
    \end{subfigure}
    \hspace*{-0.2cm}
    \begin{subfigure}{0.23\textwidth}
        \includegraphics[width=\textwidth]{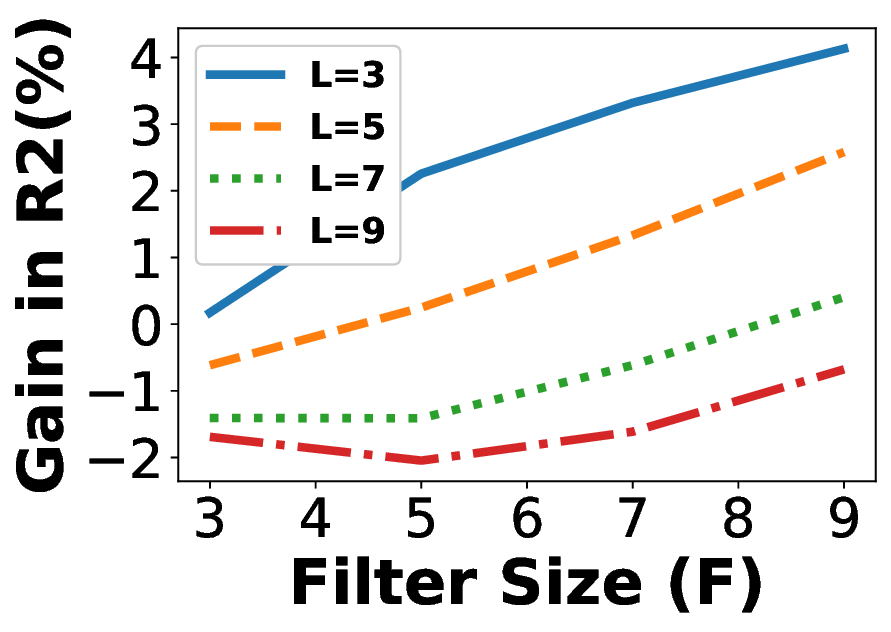}
        \caption{{Dataset: MNWILD-TNSA.}}
        \label{fig:r2vsfilter_TNSA}
    \end{subfigure}   
    \caption{{Gain in $R^2$ vs. Filter Window Size, for Varying $L$; Gain in $R^2$ is the improvement in $R^2$ score of \ac{KFTP} over standalone \ac{MLR}}.\vspace{-0.6cm}}
    \label{fig:r2vsfilter}
\end{figure}
\indent As discussed in \sectionlabel\ref{sec:Methodology}, the state equation (\ref{eq:LSE}) of our proposed KFTP algorithm is an \ac{MLR} model, which takes the present network features as input to predict the future values of network throughput. Few works~\cite{Nasri2019,Hameed2021} in the literature have suggested the use of MLR for predicting the future throughput. However, the key difference between simply using an MLR and our current work is that we have accounted for the measurement noise and the prediction error associated with  the state equation (\ref{eq:LSE}) and have recommended the correction of the same. Therefore, KFTP provides a better prediction of future throughput, even when the noise associated with the measurement set up is high. In general, measurement noise can appear at the input of any throughput prediction algorithm and perturb the performance of the quality of prediction.\\
\indent To demonstrate the robustness of our KFTP algorithm against input noise, we have performed a parametric analysis by varying the window sizes of the moving average filter ($F$). Higher value of $F$ correspond to increased noise power in the measured throughput values. Fig. \ref{fig:r2vsfilter} shows the percentage difference or the gain in $R^2$ score of throughput prediction obtained from KFTP over standalone MLR model for different values of the filter window $F$. While Fig. \ref{fig:r2vsfilter_ver} shows the gain in  $R^2$ score for the MN-Wild VER dataset, Fig. \ref{fig:r2vsfilter_TNSA} shows the gain in $R^2$ score for the MN-Wild TNSA dataset. It might be noted both from Figs. \ref{fig:r2vsfilter_ver} and \ref{fig:r2vsfilter_TNSA} that the gain in $R^2$ is substantially high for higher filtering size, which corresponds to a measurement set up with higher measurement noise. Therefore, unlike the other learning based throughput prediction algorithms including MLR, KFTP can be inferred to be much more robust and reliable in terms of nullifying the effects of input noise and is capable of predicting throughput reliably for noisy measurements.\\
\indent It may, therefore, be inferred that KFTP is able to deliver accurate throughput predictions within short inferencing times even in a noisy environment. It is also capable of quick retraining when the location of a mobile user is changed. Short training and prediction times have the potential of saving the amount of energy consumption of the end-user device. The simplistic training of KFTP entails a solution of the ordinary least square. Therefore, \ac{KFTP} seems to be much less expensive than the baseline algorithms in terms of battery drainage. Implementation of the throughput prediction algorithms in a mobile phone will render further insights into the above hypothesis and is being considered as an immediate extension of the current work. 

\section{Application: Video Streaming as Case Study}\label{sec:VS}
This section discusses the utility of the proposed \ac{KFTP} in improving the performance of network-aware applications, such as \ac{ABR} video streaming. We have considered two such applications - 1) Video-on-Demand (VoD) streaming and 2) Live video streaming. The details follow.
\subsection{VoD Streaming}\label{sec:VoDstreaming}
\indent \ac{VoD} is a system of  streaming pre-recorded media over the Internet. Such videos are streamed online using the \ac{DASH}~\cite{Mondal2020} protocol, in which a target video is broken into chunks of fixed and equal playback duration and then stored at a \ac{CDN} server at different bitrates. The \ac{DASH} video client at a \ac{UE}  uses an \ac{ABR} video streaming algorithm (at the ABR server) which aims to maximize the users' Quality of Experience (QoE) by selecting optimal bitrates for the future video chunks. To select the bitrates, the \ac{ABR} algorithm estimates the current playing conditions, such as the network bandwidth and the playback buffer length. An inaccurate estimate can inadvertently worsen the user's \ac{QoE}. For example, an optimistic estimate in the face of poor network conditions can result in fetching of chunks at high bitrates eventually slowing or \textit{stalling} the video playback, a phenomenon called \textit{rebuffering}. On the other hand, if the estimate is low even when the available network bandwidth is high, then the video will be rendered at a low bitrate,  affecting the \ac{QoE} negatively. Besides, incessant video quality fluctuations between two successive chunks also degrade the user's \ac{QoE}.\\
\indent To define the \ac{QoE} of a user mathematically, let us consider a \ac{CDN} server which stores $\mathcal{N}_T$ tracks of a video. Each track has the same video with different qualities (bitrates). Let there be $\mathcal{N}_v$ video chunks in each track, and $\zeta_i$ and $R_i$ be the length (in seconds) and the bitrate (in bps) of the $i^{th}$ chunk, respectively. Let $C_i$ (in bps) and $\mathcal{B}_i$ (in seconds) be the available network throughput and buffer level on the client side, respectively, while downloading the $i^{th}$ chunk. The \ac{QoE} is given by  \cite{Yin2015}:
\begin{equation} \label{eq:QoE}
    QOE= \sum_{i=1}^{\mathcal{N}_v}R_i - \lambda \sum_{i=1}^{\mathcal{N}_v-1}\left|R_{i+1}-R_i \right| - \mu \sum_{i=1}^{\mathcal{N}_v} \textbf{1}(\frac{\zeta_i R_i}{C_i}-\mathcal{B}_i).
\end{equation}
\indent Here, the first term represents the bitrate (in bps), the second term represents the  bitrate fluctuation (in bps), and the last term represents the stall time (in seconds). The indicator function $\textbf{1}(z)$ returns $z$ for $z>=0$, and $0$ for $z<0$. 
Thus, it is evident that the \ac{QoE} of a user can be increased by -- \begin{enumerate*}
    \item rendering the video at high bitrates,
    \item reducing the fluctuation in the bitrates of two successive video chunks, and
    \item reducing the stall or rebuffering time.
\end{enumerate*} 
 \ac{QoE} maximization is, therefore, a multiobjective optimization problem with conflicting design objectives which is addressed through the choice of bitrate selection $R_i$. The coefficients $\lambda$ and $\mu$ determine the effect of bitrate fluctuation and the rebuffering time on the \ac{QoE}. We next discuss the \ac{ABR} video streaming algorithms.
 \subsubsection{ABR video streaming}
 \begin{figure}
     \centering
     \includegraphics[width = 0.4\textwidth, trim = {0cm 1cm 0cm 0cm}]{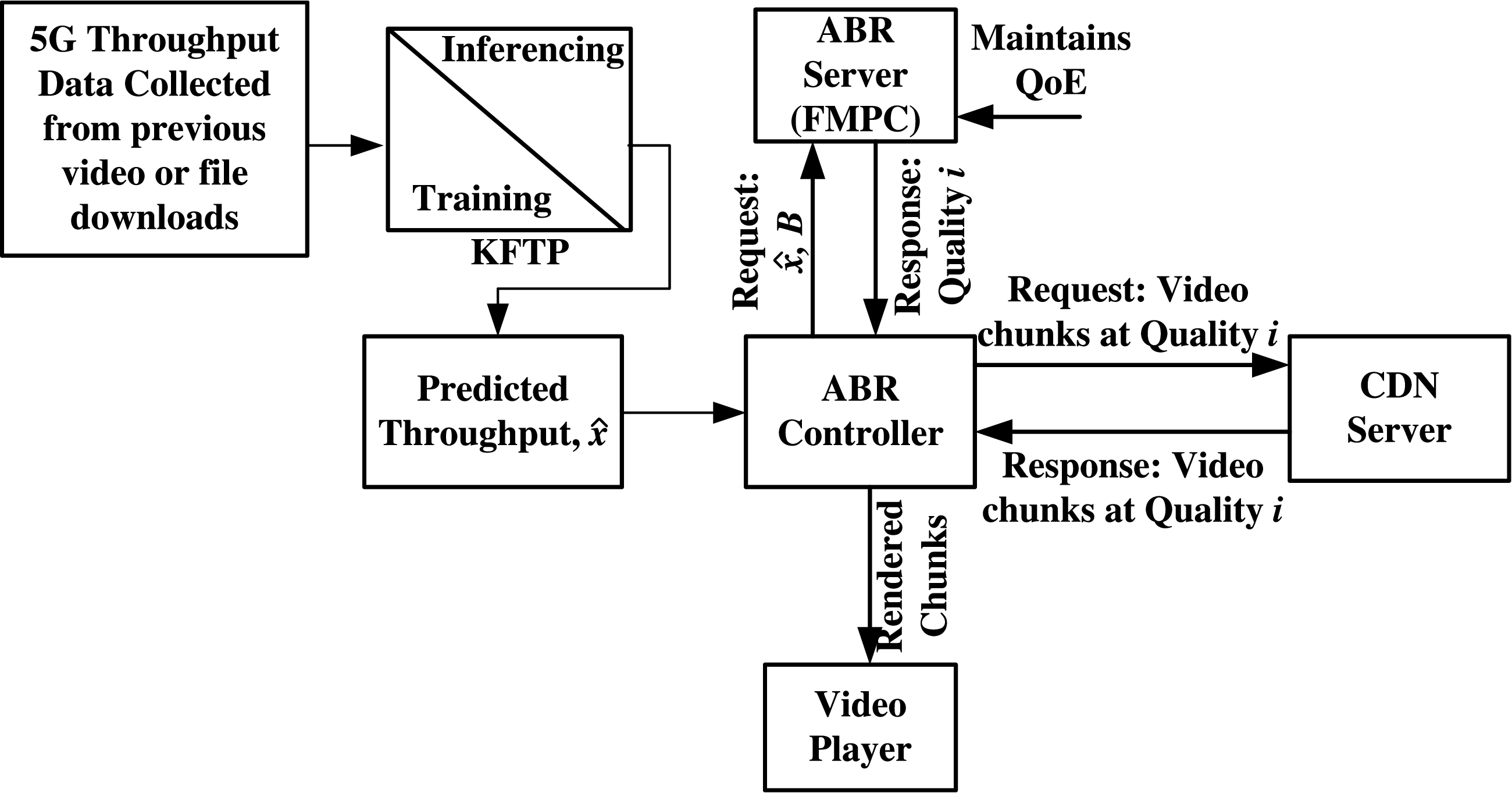}
     \caption{{A Roadmap on how KFTP can be used with ABR streaming for improving QoE in 5G.}}
     \label{fig:roadmap_kftp}
 \end{figure}
Popular state-of-art \ac{ABR} video streaming algorithms can be classified as follows:
 \begin{enumerate}
     \item \textbf{Rate based algorithms} - which use the information of the past  chunks to predict the future  throughput and optimize the bitrate of the future chunks, for example, FESTIVE~\cite{Jiang2014},
     \item \textbf{Control Theoretic} - which solve an optimization problem to decide the bitrate of the future video chunks, example, Fast MPC (\ac{FMPC})~\cite{Yin2015}, \ac{RMPC}~\cite{Yin2015},
     \item \textbf{Buffer based algorithms} - which use only the playback buffer status information to decide chunk bitrates, for example, BOLA~\cite{Spiteri2016},
     \item \textbf{Learning based algorithms} - which use neural networks to make optimal bitrate decisions for maximizing the \ac{QoE}, for example, Pensieve~\cite{mao2017neural}, Oboe~\cite{Akhtar2018}.
 \end{enumerate}
  \indent Of these, the \textbf{\ac{FMPC}} algorithm selects the bitrates $R_i$ by considering the future throughput over $N_F$ future steps. The flow of \ac{FMPC} may be enumerated as follows:
  \begin{enumerate}
      \item At any iteration `$i$' the player maintains a moving horizon for $N_F$ chunks into the future, i.e., from chunk $i$ to $i+N_F-1$. It then predicts the throughput $\hat{C}_{[t_i, t_{i+N_F}]}$. Authors in \cite{Yin2015} use a lookahead horizon of $h=5$ chunks and the throughput is predicted as the harmonic mean of the previous five chunks. \textit{In our work, we are going to replace the harmonic mean throughput predictor of~\cite{Yin2015} with our proposed \ac{KFTP}}. {Fig. \ref{fig:roadmap_kftp} shows how KFTP can be used with the \ac{FMPC} algorithm hosted at the ABR server to improve the QoE of VoD streaming in 5G.}
      \item \ac{FMPC} then takes the predicted throughput $\hat{C}_{[t_i, t_{i+N_F}]}$, the buffer level $\mathcal{B}_i$ and the previous bitrate $R_{i-1}$ to select the optimal chunk bitrates $R_i$ at the chunk boundaries, in order to maximize the \ac{QoE} of (\ref{eq:QoE}). To optimize the \ac{QoE}, \ac{FMPC} solves the optimization problem in~\cite[Fig. 3]{Yin2015}. In the steady state, only the chunk bitrates are predicted, i.e., $R_i=f_{mpc}(R_{i-1}, \mathcal{B}_i, \hat{C}_{[t_i, t_{i+N_F}]})$. In the start up phase, \ac{FMPC} also optimizes the startup time $T_s$. 
      \item \ac{FMPC} then downloads chunk $i$ at bitrate $R_i$ and shifts the look ahead horizon to the next $N_F$ chunks. 
  \end{enumerate} 
  \begin{figure*}[t]
     \centering
   \begin{subfigure}{0.23\textwidth}
         \centering
         \includegraphics[width = \textwidth]{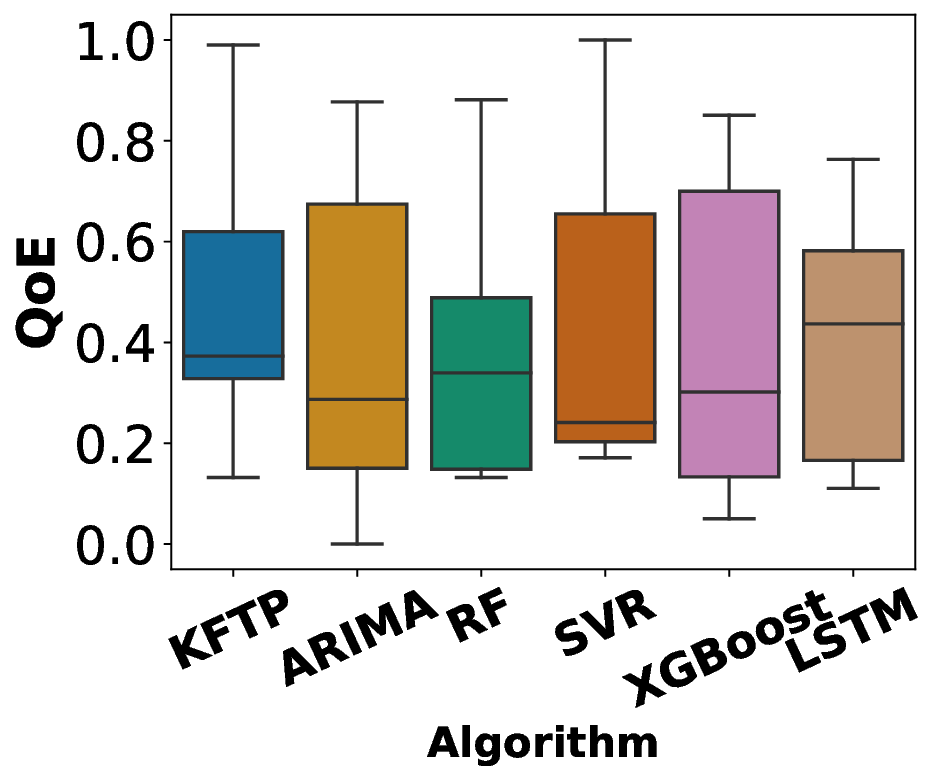}
     \caption{{ QoE (Normalized).} }
        \label{fig:vs-qoe}
     \end{subfigure}
     \hspace{-0.2cm}
    \begin{subfigure}{0.23\textwidth}
        \centering
        \includegraphics[width = \textwidth]{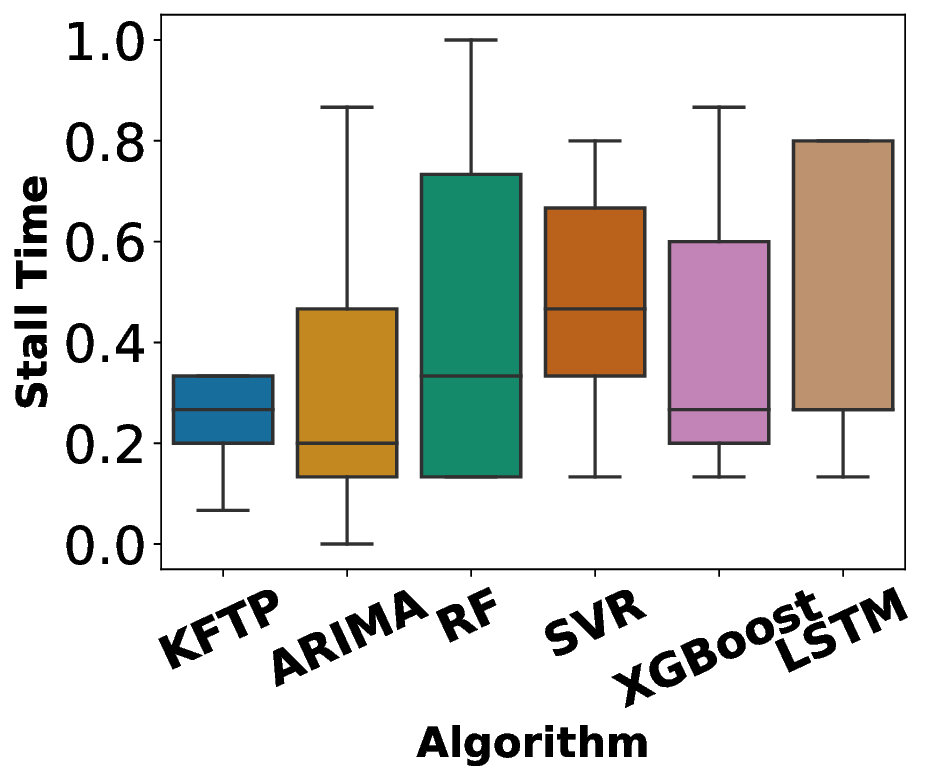} 
        \caption{{Stall Time (Normalized).}}
        \label{fig:vs-stall}
    \end{subfigure}
\hspace*{-0.2cm}
     \begin{subfigure}{0.23\textwidth}
         \centering       \includegraphics[width = \textwidth]{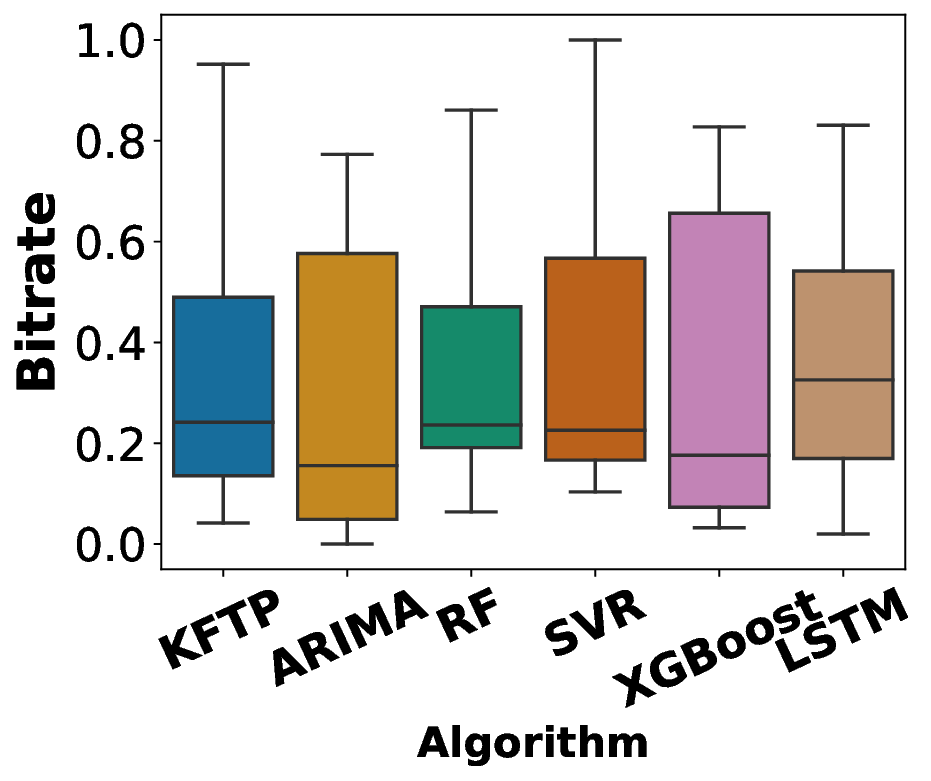}
        \caption{{Bitrate (Normalized)}.}
        \label{fig:vs-bitrate}
     \end{subfigure}
     \hspace*{-0.2cm}
     \begin{subfigure}{0.23\textwidth}
         \centering
         \includegraphics[width = \textwidth]{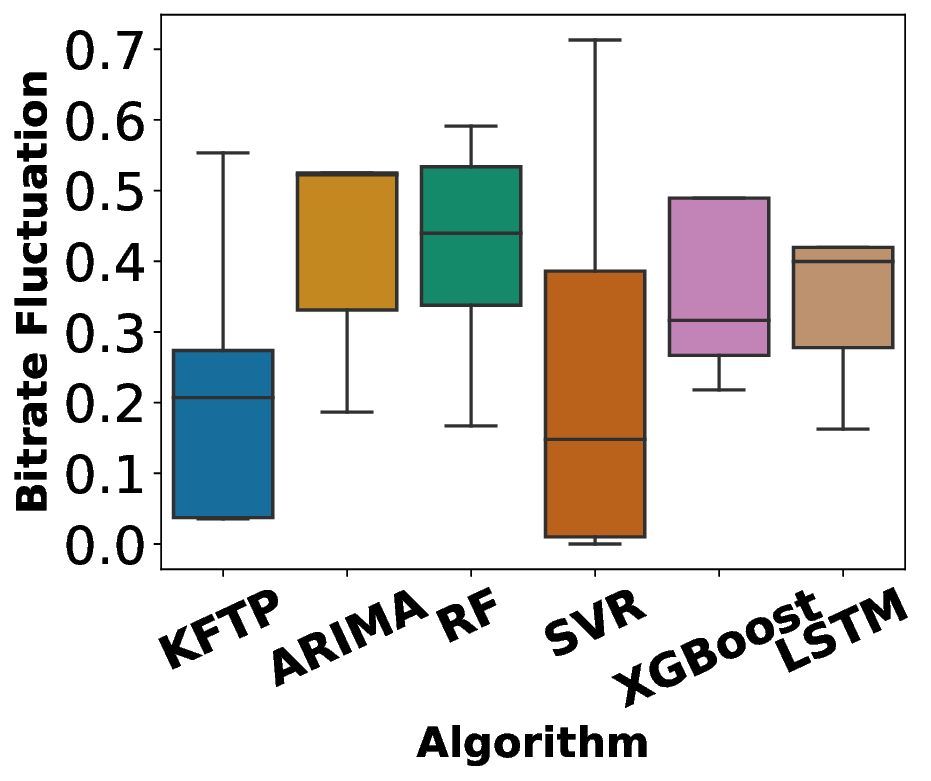}
        \caption{{Bitrate Fluctuation (Normalized).}}
        \label{fig:vs-bitrate-difference}
     \end{subfigure}
    \centering\caption{{Comparison of the Different Parameters Constituting \ac{QoE} in VoD streaming, for Different Throughput Predictors, \predlength{5} seconds, \filterwindow{3} samples Throughput Dataset: MNWILD-TNSA.}}
    \label{fig:vs-qoe-parameters}
\end{figure*}A state in the \ac{FMPC} algorithm depends on the current buffer level, the previous bitrates and the future predicted throughput. As a result, the state space can be huge necessitating large memory storage and a large computational overhead. Hence, to ensure tractability, authors in \cite{Yin2015} have compacted the state space by considering 100 bins of buffer level and 100 bins of predicted throughput to achieve near-optimal performance.
\subsubsection{Testbed and Simulation Setup}  To demonstrate the \ac{QoE} performance improvement offered by \ac{KFTP}, we have replaced the harmonic mean based throughput prediction of \ac{FMPC} with each of the throughput prediction algorithms discussed in \sectionlabel\ref{sec:Results}, including our proposed \ac{KFTP}. Our testbed has a video client-server setup as in~\cite{Narayanan2021}\footnote{https://github.com/SIGCOMM21-5G/artifact/tree/main/Video-Streaming} , in which the video server is hosted in an Apache server \cite{Narayanan2021} while the client runs a dash.js video player in a web browser. The measured and predicted throughput traces used in our experiment correspond to the  TMobile, NSA+LTE (MNWILD-TNSA) dataset, which supports mmWave communication. The video used is of 4K resolution~\cite{VideoYoutube2016}, consisting of 6 separate tracks. As in~\cite{Narayanan2021}, we have scaled the bitrates of these six tracks to make them commensurate with the high throughput of the MNWILD-TNSA data trace. The different bit rates are  -- 20 Mbps, 40 Mbps, 60 Mbps, 80 Mbps, 110 Mbps, 160 Mbps. The length of this specific video file is 158 seconds, which is broken down into 157 chunks. The future chunk length $N_F$ is set to 5, which approximates to $5 \times \frac{158}{157} \simeq $  5 seconds. The values of $\lambda$ and $\mu$ of (\ref{eq:QoE}) are 1 and 160 Mbps, respectively.
\subsubsection{Observations}\label{sec:VoDs-observations} Fig. \ref{fig:vs-qoe} shows the normalized \ac{QoE} offered by \ac{FMPC} in conjunction with the proposed \ac{KFTP} and the baseline prediction algorithms, averaged over five MNWILD-TNSA  traces for a time lead of \predlength{5} seconds and a filtering window of \filterwindow{3} samples. It is seen that the 50$^\text{th}$ percentile point of the \ac{QoE} offered by \ac{KFTP} is higher than all the other throughput prediction algorithms except \ac{LSTM}. The latter, however, has a high training and prediction time as may be observed from Fig. \ref{fig:complexity}. 
 \begin{figure}
    \centering
    \includegraphics[width = 0.45\textwidth, trim = {0cm 0.0cm 0cm 0cm}, clip]{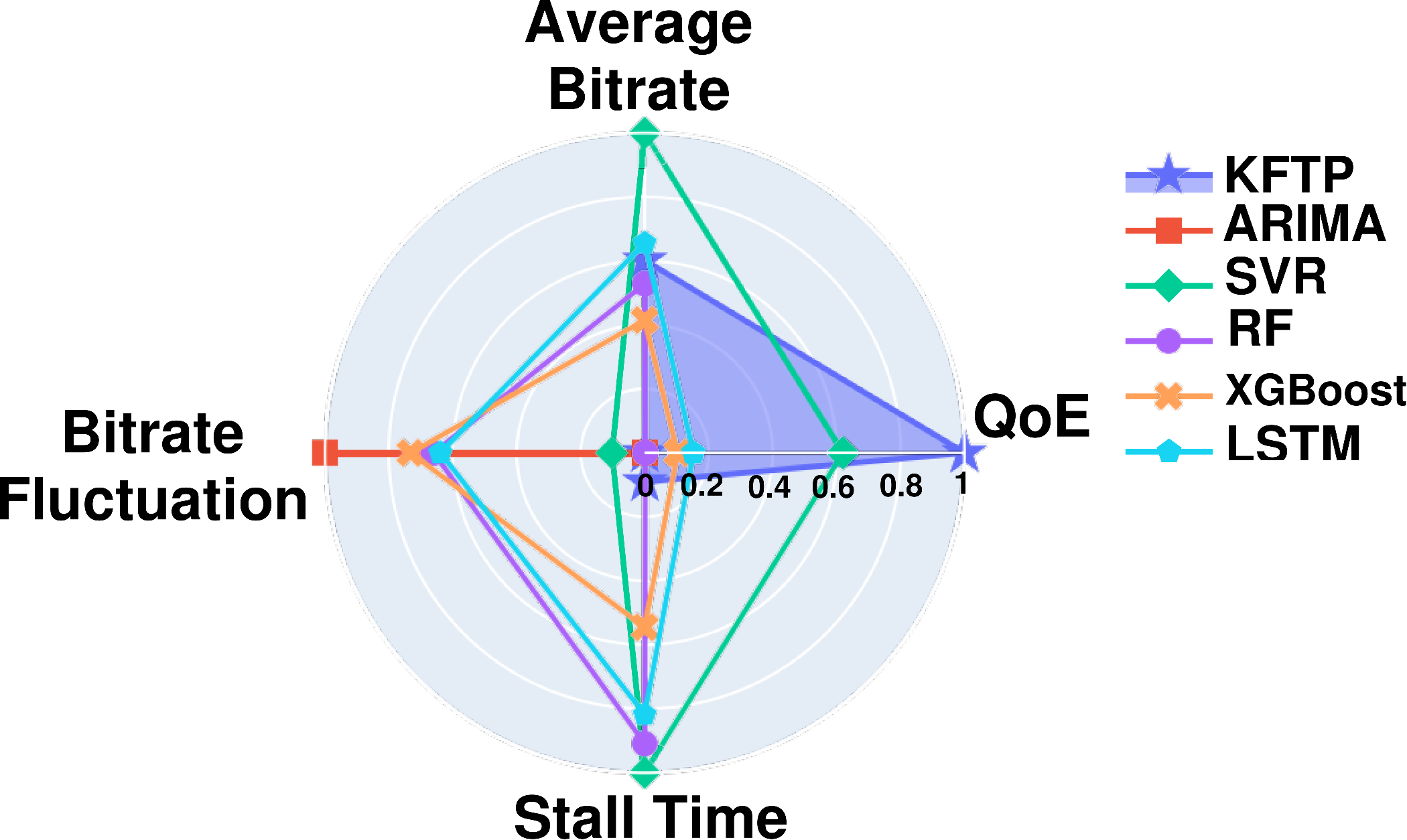} 
    \caption{{Spider plot of Normalized  Bitrate, Bitrate Fluctuation, Stall Time, and \ac{QoE} of  \textit{FMPC}, for \predlength{5} seconds, \filterwindow{7} samples, MNWILD-TNSA dataset. (The value 1 on the \ac{QoE} axis is the maximum value of normalized \ac{QoE}.)}}
    \label{fig:spider-plot}
\end{figure}
{Figs. \ref{fig:vs-stall}, \ref{fig:vs-bitrate}, \ref{fig:vs-bitrate-difference} show the normalized values of average stall time, average bitrate, and average bitrate variation corresponding to Fig. \ref{fig:vs-qoe}.} A consolidated view in the form of a spider plot is also provided in Fig. \ref{fig:spider-plot}.  It may be noted that the high throughput of \ac{KFTP} is primarily due to the small stall time, which incurs the highest penalty among all the components of \ac{QoE}. The reduced stall time can be attributed to the high $R^2$ score of \ac{KFTP}. The average \ac{QoE} (not normalized) of \ac{FMPC} observed for \ac{KFTP}, \ac{ARIMA}, \ac{RF}, \ac{SVR}, \ac{XGBoost}, and \ac{LSTM} algorithms are -- 83.71, 83.02, 83.02, 83.45, 83.09, 83.12, respectively, for \predlength{5} seconds and \filterwindow{3} samples. {The \ac{QoE} offered by \ac{FMPC} when using \ac{KFTP} is higher than or at par with the other existing throughput prediction algorithms, while having shorter training and prediction times. The reduced prediction time makes KFTP suitable for use in 5G mmWave as the channel condition of the latter is highly dynamic.}
\subsection{Live Streaming}
Live streaming is another revenue-generating application which can benefit significantly from the improved accuracy of throughput prediction.  As live streaming videos are segmented, encoded and streamed in real-time, they are critically sensitive to end-to-end playback latency. However, variations in the underlying network conditions may delay the delivery of video segments, which can violate their playback deadlines. Consequently, there is an increase in the number of \textit{video freeze} or \textit{video stall} events, which in turn increases the \textit{end-to-end playback latency} and thereby reduces the user's \ac{QoE}. Live video streaming algorithms aim to maximize the \ac{QoE} through an adaptive selection of the video segment bitrates while maintaining a low end-to-end playback latency. \\ 
\indent  The \ac{QoE} of live video users has been derived in detail in~\cite{monkeysun2019}. 
It has been assumed in~\cite{monkeysun2019} that the client downloads the segments sequentially in real-time, i.e., it requests segment $i$ from the server only if segment $(i-1)$ has been completely downloaded. Additionally, segment $i$ can only be downloaded if it has been completely encoded at the server. It is further assumed that all  segments are of the same duration, i.e., of $\zeta=1$ second. Let $R_i$ (in bps) represent the video rate of segment $i$, and $C_i$ (in bps) and $\mathcal{B}_i$ (in seconds) represent the available network throughput and client-side buffer level, respectively, while downloading segment $i$. The total download time of segment $i$ is, thus, $t^D_{i}=\frac{R_i \zeta}{C_i}+\Delta_i$. Here $\Delta_i$ accounts for the overhead associated with -- a) the round trip time delay and b) the idle time which arises when the requested segment is not ready for download. Any instance of freezing or stall at the client, with a stall time of $t^\mathrm{stall}_{i}=\textbf{1}(t^\mathrm{D}_{i}-\mathcal{B}_i)$, results in an increment in the latency $l_{i-1}$ by $t^\mathrm{stall}_{i}$. 
Whenever $l_i$ crosses a maximum permissible latency $l_{max}$ second, the client drops $\eta_i$ segments to catch up with the live feed in real time. The modified \ac{QoE} expression for live streaming~\cite{monkeysun2019} is, thus, defined as:
 \begin{align}\label{eq:qoelive}
     QoE_\mathrm{live}  = & \sum_{i=1}^{\mathcal{N}_v}w_1 Q(R_i) - w_2 t^\mathrm{stall}_{i} \nonumber\\ 
     &-w_3\left|Q(R_{i+1})-Q(R_{i}) \right|-w_4\phi(l_i)- w_5\eta_i,
 \end{align}
where $Q(r)=log(r/R_{min})$ is the perceptible video quality. $\phi(l_i)$, which represents the playback latency. It is a logistic growth function defined in \cite{monkeysun2019} as $\frac{1}{1+e^{\omega-l_i}}-\frac{1}{1+e^{\omega}}$. While $\omega$ determines the latency sensitivity range, the coefficients $w_1, \ w_2, \ w_3,\ w_4, \ w_5$ in (\ref{eq:qoelive}) determine the users' sensitivity to the different \ac{QoE} components. These values may be tuned to maximize the \ac{QoE}.  Moreover, the live streaming algorithms of~\cite{monkeysun2019} follow a $(\alpha, \beta)$ strategy, i.e., when a user joins the live stream, it can request to download  a maximum of $\alpha$  already encoded segments only, and the live stream starts when at least $\beta$ segments have been downloaded. We next discuss the live streaming algorithms. 
\begin{figure*}
     \centering
   \begin{subfigure}{0.3\textwidth}
         \centering
         \includegraphics[width = \textwidth]{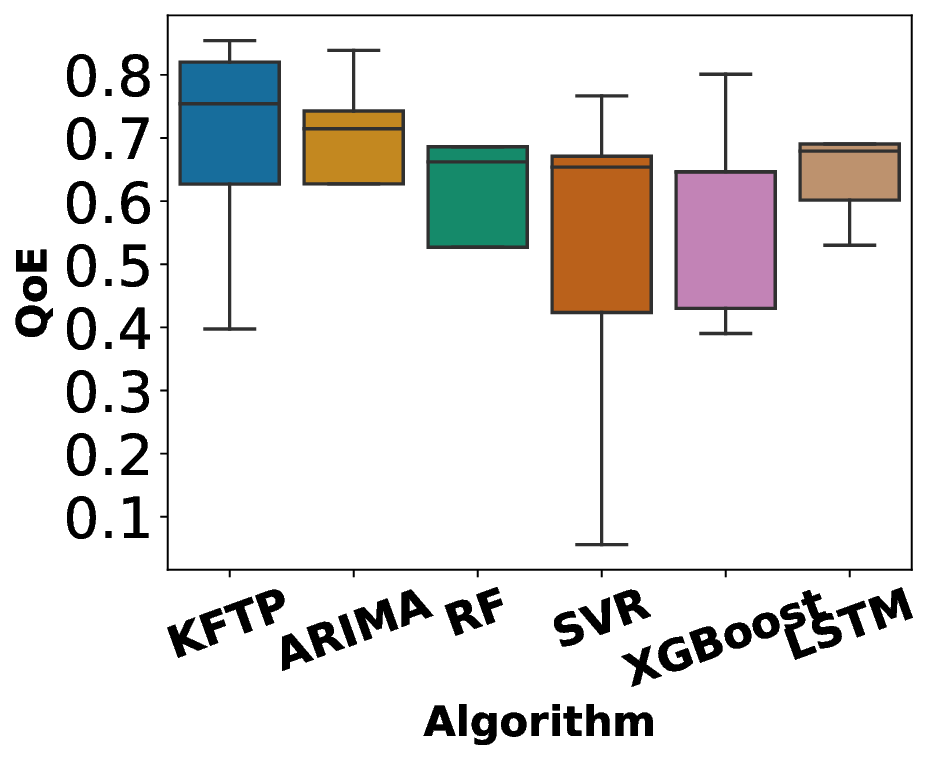}
     \caption{{Normalized QoE.} }
        \label{fig:ls-qoe}
     \end{subfigure}
    \begin{subfigure}{0.3\textwidth}
        \centering
        \includegraphics[width = \textwidth]{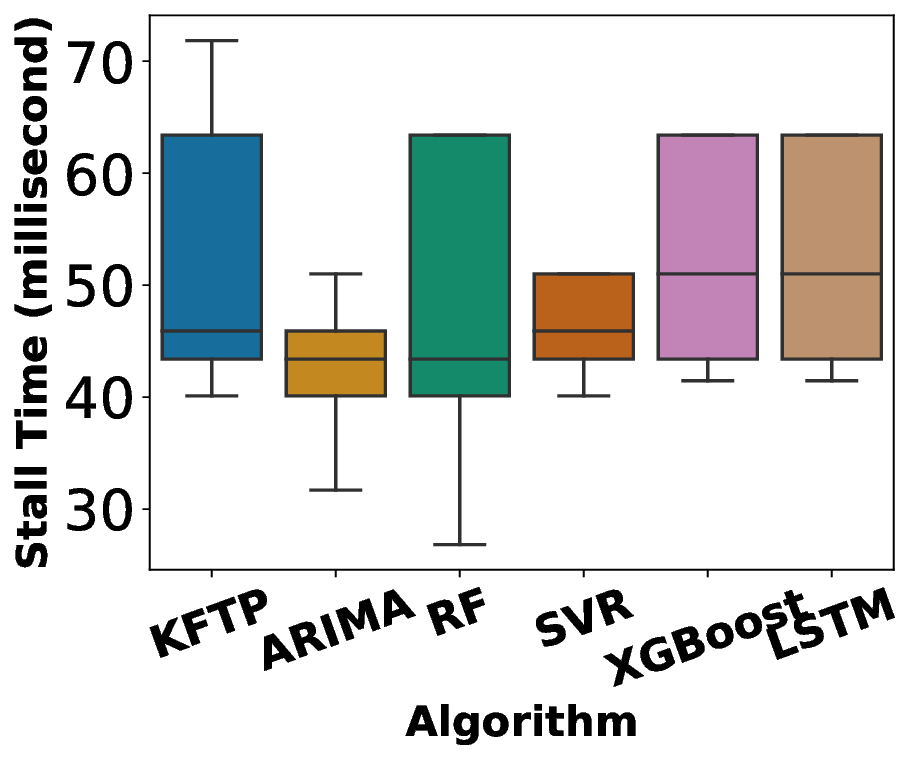} 
        \caption{{Stall Time (millisecond).}}
        \label{fig:ls-stall}
    \end{subfigure}
     \begin{subfigure}{0.3\textwidth}
         \centering       \includegraphics[width = \textwidth]{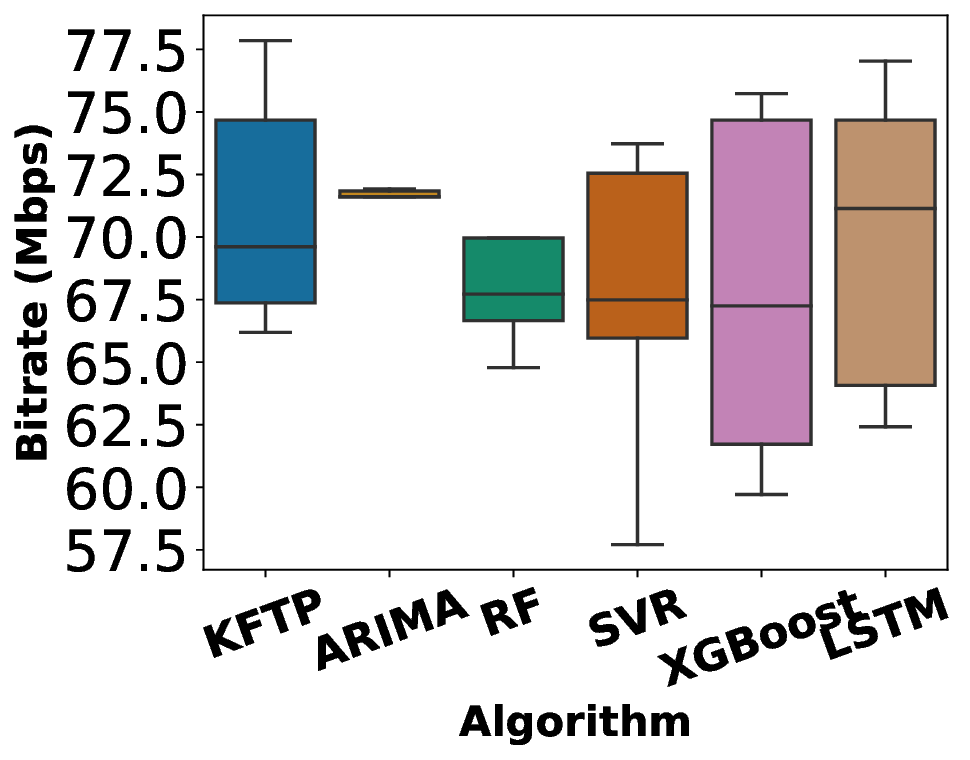}
        \caption{{Bitrate (Mbps).}}
        \label{fig:ls-bitrate}
     \end{subfigure}
      \begin{subfigure}{0.3\textwidth}
         \centering
         \includegraphics[width = \textwidth]{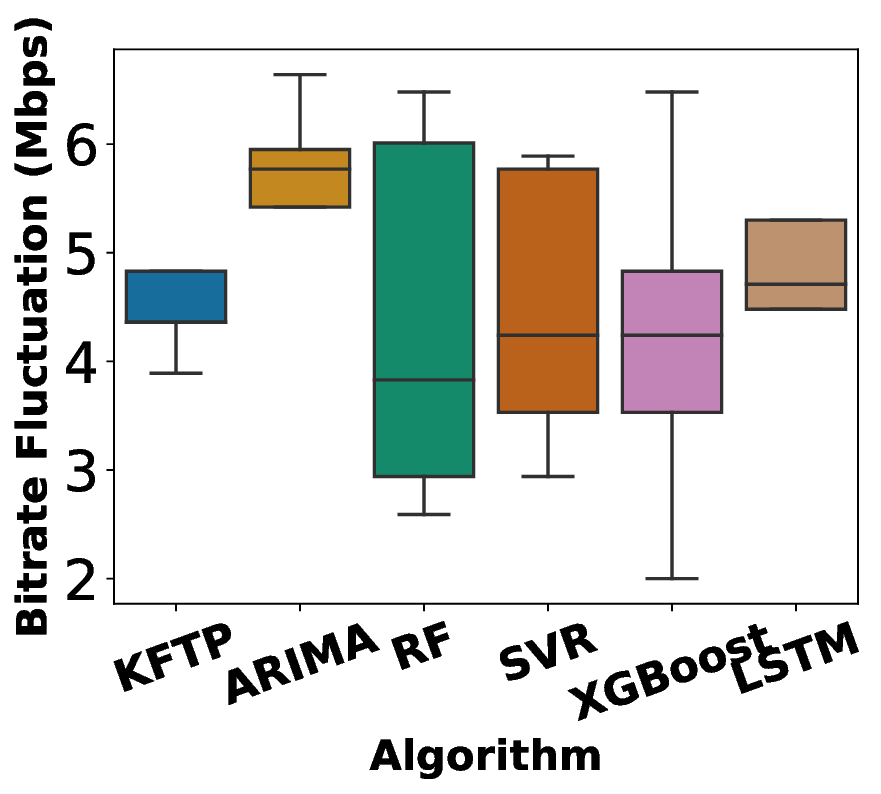}
        \caption{{Bitrate Fluctuation (Mbps).}}
        \label{fig:ls-bitrate-difference}
     \end{subfigure}
     \begin{subfigure}{0.3\textwidth}
         \centering
         \includegraphics[width = \textwidth]{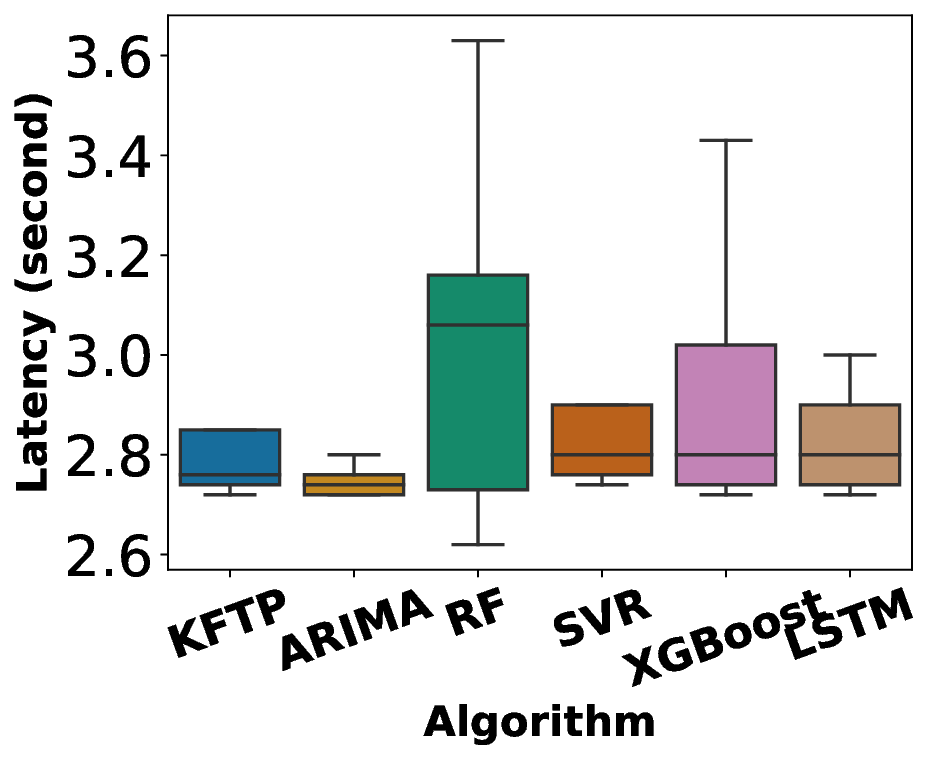}
        \caption{{Latency (second).}}
        \label{fig:ls-latency}
     \end{subfigure}
    \centering\caption{{Comparison of the Different Parameters Constituting \ac{QoE} in Live Streaming, for Different Throughput Predictors, \predlength{5} seconds, Dataset: MNWILD-TNSA.}}
    \label{fig:ls-qoe-parameters}
     \end{figure*}
     \begin{figure}[t]
\centering
    \includegraphics[width = 0.45\textwidth, trim = {0cm 1cm 0cm 2cm}]{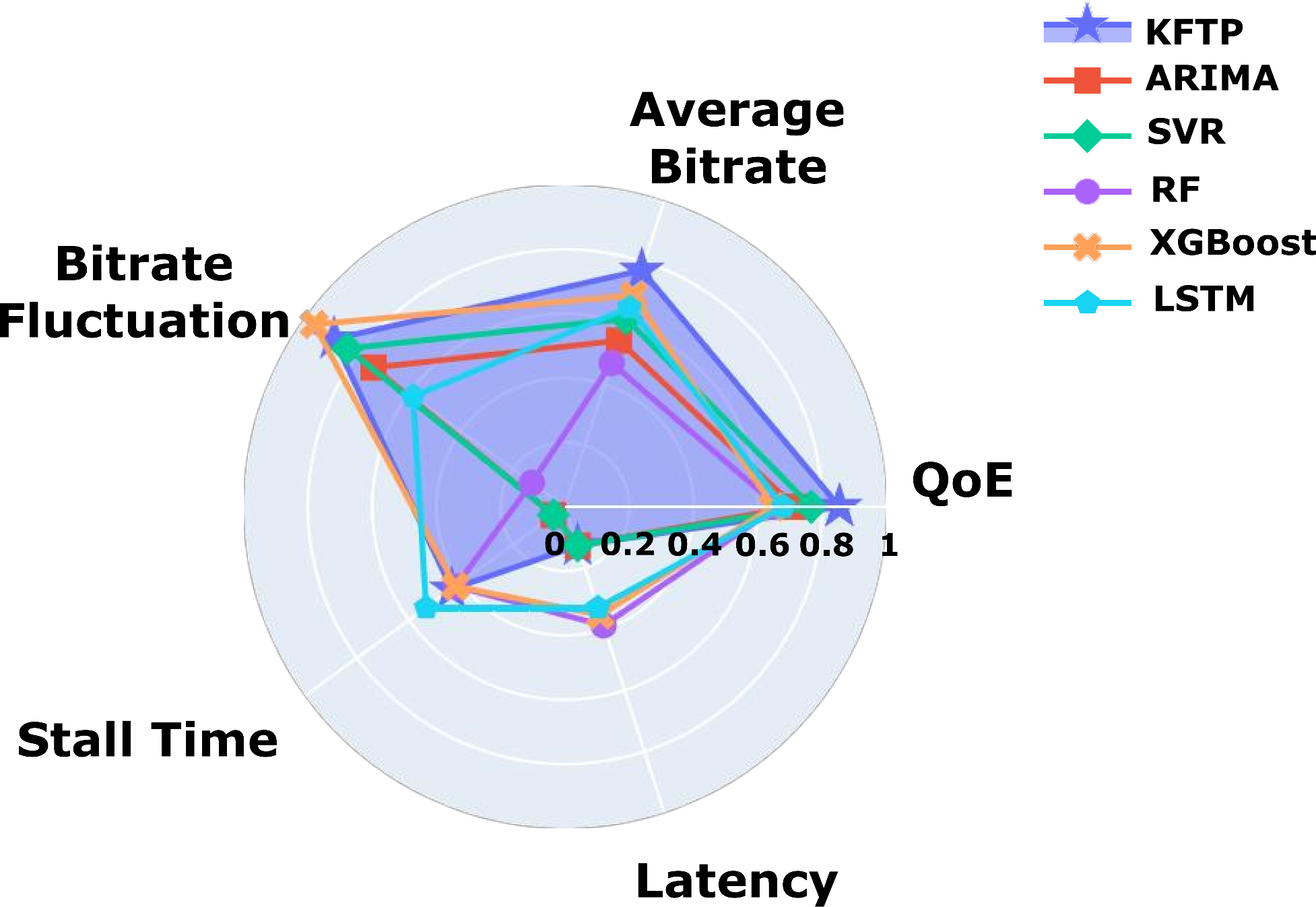} 
    \caption{{Spider plot of Normalized  Bitrate, Bitrate Fluctuation, Stall Time, and \ac{QoE} of $MPC_{live}$ for Different Throughput Predictors, \predlength{5} seconds, MNWILD-TNSA dataset.}}
     \label{fig:spider-plot-ls}
\end{figure}
\subsubsection{MPC for Live streaming Videos}
Authors in \cite{monkeysun2019} have proposed a \acf{MPC} based live video streaming algorithm, \textit{MPC-Live}~\cite[Algorithm 2]{monkeysun2019}. It captures the interaction between video rates, video freeze, and end-to-end playback latency to maximize the user's \ac{QoE} while constraining the playback latency to remain below a predefined threshold.  \textit{MPC-Live} needs the network throughput $C_i$ and the round trip time $\mathrm{rtt}_i$  to select the optimal bitrate $R_i$ for segment-$i$. However, as these quantities are not available apriori, \textit{MPC-Live} uses the predicted network throughput and the predicted round trip time over a moving horizon of $N_F$ segments to select $R_i$. The predicted throughput of the $i^{th}$ segment is the harmonic mean of the previous $i-N_F+1$ chunks. Once $R_i$ is selected, \textit{MPC-Live} downloads  segment-$i$ at $R_i$,  and then shifts the horizon to the next $N_F$ chunks. \\
\indent  In DASH, segments can be downloaded only when they are completely encoded, thereby introducing an additional delay of one segment. To counteract this issue,~\cite{monkeysun2019} has proposed a second algorithm called \textit{MPC-Chunk}, in which a video segment is split into chunks of smaller duration according to~\cite{CMAF2020}. In this, chunks can be sent to the client as and when they are encoded. As a result, the playback latency can be significantly reduced. In our work, we have used the \textit{MPC-Chunk} algorithm for demonstrating the usefulness of \ac{KFTP} in improving the \ac{QoE} of live streaming users. 
\subsubsection{Simulation Setup} 
In this paper, we have evaluated the performance of the \textit{MPC-Chunk} algorithm, using the live-video streaming simulator\footnote{https://github.com/monkeysun555/benchmark\_mpc} of~\cite{monkeysun2019}. The existing simulator can support 4G throughput traces only. So, we have modified it to work with predicted 5G throughput traces. We have also included a bitrate up-scaling feature as in Section \ref{sec:VoDstreaming} so that the supported bitrates are commensurate with the high network throughput of 5G. Thus, in our work, the live video is streamed in the following resolutions  -- 240p (300 Kbps), 360p (500 Kbps), 480p (1000 Kbps), 720p (2000 Kbps), 1080p (3000 Kbps), and 1440p (6000 Kbps).  The live video of 150 seconds is encoded into 1-second segments. Each segment is further subdivided into 5 chunks, each of 200 milliseconds. The playback threshold at the client and the maximum latency threshold $l_{max}$ are set to 2 seconds and 5 seconds, respectively.   The values of $\alpha$ and $\beta$ are set as $\alpha=3$ and $\beta=2$, i.e., a user joining the stream can request at most three already encoded video segments, and the streaming starts when it has downloaded at least two segments. The values of the coefficients $w_1, \ w_2, \ w_3,\ w_4, \ w_5$ of (\ref{eq:qoelive}) are 0.2, 6.0, 1.0, 0.8, and 1.2, respectively, implying that an increase in the video stall time incurs the highest penalty.\\
\indent In order to demonstrate how the proposed \ac{KFTP} algorithm improves the live-streaming \ac{QoE}, we have replaced the harmonic mean throughput predictor of \textit{MPC-Live} with  \ac{KFTP} and other baseline throughput prediction algorithms. The baseline throughput prediction algorithms considered in this section are the same as in Section \ref{sec:VoDstreaming}, i.e., a) ARIMA, b) Random Forest (RF), c) Support Vector Regressor (SVR), d) Extreme Gradient Boost (XGBoost), and e) Long Short Term memory (LSTM).
\subsubsection{Observations}
Fig. \ref{fig:ls-qoe} shows the normalized \ac{QoE} of \textit{MPC-Chunk}, averaged over five MNWILD-TNSA throughput traces, for a time lead of \predlength{5} seconds and a filtering window of \filterwindow{3} samples. It is observed that \ac{QoE} of \ac{KFTP} is considerably higher than the baseline algorithms. Of the baseline algorithms, LSTM delivers the best performance with a normalized median \ac{QoE} of 0.679. However, the normalized median \ac{QoE} of \ac{KFTP} is 0.754, which is almost 11\% higher than \ac{LSTM}. This is due to the fact that although the stall time (Fig. \ref{fig:ls-stall}) and the bitrate (Fig. \ref{fig:ls-bitrate}) of \ac{KFTP} is at par with \ac{LSTM}, its bitrate variation (Fig. \ref{fig:ls-bitrate-difference}) and playback latency (Fig. \ref{fig:ls-latency}) is comparatively lower. Fig. \ref{fig:spider-plot-ls} gives a consolidated view of the normalized \ac{QoE} and its constituent parameters. It corroborates that the high \ac{QoE} of \ac{KFTP} is due to its high average quality, low latency, and stall time as compared to the baseline algorithms. It should be noted that the playback latency threshold of five seconds was never violated during the simulations with the 5G traces. As a result, no segment has been dropped. It may, therefore, be said that the proposed \ac{KFTP} throughput algorithm offers higher \ac{QoE} to live video streaming users than the existing algorithms while being computationally less intensive than them.

%% file: Sections/Conclusion.tex
\section{Conclusion}
\label{sec:Conclusion}
\acresetall
In this work, we have proposed KFTP a low computationally complex approach for predicting the throughput of 5G enhanced Mobile Broadband (eMBB) mmWave networks. Existing works have primarily prescribed \ac{ML} and \ac{DL} algorithms to predict 5G throughput. However, these models are likely to be highly power hungry  due to their high computational complexity  and may struggle to deliver timely predictions within the short coherence time of 5G. Moreover, these works have not accounted for the error in measuring network throughput, which may lead to unreliable predictions. Our proposed KFTP, a Kalman Filter based throughput prediction, on the other hand  leverages the prediction and correction approach of Kalman filters to predict the future throughput using a simple linear state estimation. It exploits the statistical properties of the prediction error and measurement error to obtain the optimal estimates of throughput. In the process, it uses only eight multiplications and seven addition operation, thereby being significantly less computationally intensive. We have conducted extensive experiments with seven popular 5G throughput datasets for varying lengths of the prediction window size. The results have shown that the proposed KFTP outperforms baseline algorithms and restricts the MAE to below 15\% for prediction window sizes as high as 9 seconds. In addition, it has significantly shorter training and inferencing times, making it suitable for retraining on the energy-constrained, handheld user equipments. We have also demonstrated video streaming as an application for throughput prediction. It has been observed that \ac{KFTP} delivers higher \ac{QoE} for both video-on-demand and live streaming, by reducing the stall time and the playback latency, respectively. 
Thus, it seems that \ac{KFTP} can be used as a low computationally complex throughput prediction algorithm for delivering energy and time efficient throughput estimates.

%% file: acronyms.tex
\begin{acronym}
	\acro{2G}{2$^\mathrm{nd}$ Generation}
	\acro{3G}{3$^\mathrm{rd}$ Generation}
	\acro{4G}{4$^\mathrm{th}$ Generation}
	\acro{5G}{5$^\mathrm{th}$ Generation}
	\acro{A3C}{Actor-Critic}
	\acro{ABR}{adaptive bitrate}
	\acro{AR}{Augmented Reality}
	\acro{ARE}{Absolute Value of Residual Error}
        \acro{ARMA}{Auto Regressive Moving Average}
	\acro{ARIMA}{Auto Regressive Integrated Moving Average}
	\acro{BLUE}{Best Linear Unbiased Estimator}
	\acro{BS}{Base Station}
	\acro{CDN}{Content Distribution Network}
 \acro{CNN}{Convolutional Neural Networks}
	\acro{CQI}{Channel Quality Index}
	\acro{D1} {Data set from \cite{Raca2020_1}}
	\acro{D2} {Data set from \cite{Narayanan2021}}
	\acro{DASH}{Dynamic Adaptive Streaming over HTTP}
	\acro{DTR}{Decision Tree Regression}
	\acro{DRX}{Discontinuous Reception}
	\acro{DL}{Deep Learning}
	\acro{DT}{Decision Trees}
        \acro{MPC}{Model Predictive Control}
        \acro{FMPC}{Fast \ac{MPC}}
	\acro{KF}{Kalman Filter}
        \acro{KFTP}{Kalman Filter based Throughput Prediction}
        \acro{KNNR}{K-Nearest Neighbours Regression}
        \acro{MA}{Moving Average}
	\acro{MAE}{Mean Absolute Error}
	\acro{mn}{Measurement Noise}
	\acro{MSE}{Mean Squared Error}
	\acro{ML}{Machine Learning}
	\acro{MLP}{Multilayer Perceptron}
	\acro{MLR}{Multiple Linear Regression}
	\acro{EDGE}{Enhanced Data Rates for \ac{GSM} Evolution.}
	\acro{eNB}{evolved NodeB}
    \acro{GBDT}{Gradient Boosting Decision Trees}
	\acro{GSM}{Global System for Mobile}
	\acro{HD}{High Definition}
	\acro{HSPA}{High Speed Packet Access}
        \acro{IoT}{Internet of Things}
        \acro{KNN}{K-Nearest Neighbours}
	\acro{LTE}{Long Term Evolution}	
	\acro{LSTM}{Long Short Term Memory}
	\acro{NR}{New Radio}
	\acro{pn}{Prediction Noise}
	\acro{HVPM}{High voltage Power Monitor}
	\acro{gNB}{gNodeB}
	\acro{GPX5} {Google Pixel 5}
    \acro{KDE}{Kernel Density Estimate}
    \acro{PDF}{probability density function}
	\acro{QoS}{Quality of Service}
	\acro{QoE}{Quality of Experience}
	\acro{RF}{Random Forest}
	\acro{RFL}{Random Forest Learning}
	\acro{RL}{Reinforcement Learning}
	\acro{RMSRE}{Root Mean Square Relative Error}
	\acro{RNN}{Recurrent Neural Network}
	\acro{RRC}{Radio Resource Control}
    \acro{RSS}{Residual Sum of Squares}
	\acro{RSSI}{Received Signal Strength Indicator}
	\acro{RSRP}{Reference Signal Received Power}
	\acro{R2}{R-squared metric}
	\acro{RSRQ}{Reference Signal Received Quality}
        \acro{SGS20U}{Samsung Galaxy S20 Ultra}
	\acro{SA}{Standalone 5G}
	\acro{SER}{Signal to Error Ratio}
	\acro{SVR}{Support Vector Regressor}
	
	\acro{NSA} {Non-standalone 5G}
 \acro{RMPC}{Robust MPC}
	\acro{SGS20U} {Samsung Galaxy S20 Ultra 5G}
    \acro{SGS10} {Samsung Galaxy S10 5G}

	\acro{SINR}{signal-to-interference-plus-noise-ratio}
	\acro{SNR}{signal-to-noise-ratio}
	\acro{TNSA}{TMobile, NSA+LTE}
	\acro{TSA}{TMobile, SA}
        \acro{THPT}{Throughput}
        \acro{EWMA}{Exponential Weighted Moving Average}
	\acro{UE}{User Equipment}
	\acro{UHD}{Ultra HD}
	\acro{Ver}{Verizon, Default}
	\acro{VoLTE}{Voice over LTE}
	\acro{VR}{Virtual Reality}
        \acro{WCDMA}{Wideband Code Division Multiple Access}
	\acro{WiFi}{Wireless Fidelity}
	\acro{XGBoost}{Extreme Gradient Boost}
	\acro{SUMO}{Simulation of Urban MObility}
	\acro{MCS}{Modulation Coding Scheme}
        \acro{HM}{Harmonic Mean}
        \acro{VoD}{video-on-demand}
\end{acronym}